\documentclass{aa} 
\usepackage{graphicx}
\usepackage{lscape}
\usepackage{longtable}
\usepackage{arydshln}
\usepackage{multirow}
\usepackage{amsmath}
\usepackage{enumitem}
\usepackage[version=4]{mhchem}
\usepackage{txfonts}
\usepackage{natbib}
\usepackage[htt]{hyphenat}

\bibpunct{(}{)}{;}{a}{}{,} 

\begin{document}

\title{AB Aur, a Rosetta stone for studies of planet formation (II): $\rm H_2S$ detection and sulfur budget} 
   \author{P. Rivi\`ere-Marichalar\inst{1}, A. Fuente\inst{1}, G. Esplugues\inst{1}, V. Wakelam\inst{2}, R. le Gal\inst{3,4}, C. Baruteau\inst{5}, A. Ribas\inst{6}, E. Mac\'ias\inst{6,7,8}, R. Neri\inst{9}, D. Navarro-Almaida\inst{1}}
 \institute{Observatorio Astron\'omico Nacional (OAN,IGN), Calle Alfonso XII, 3. 28014 Madrid, Spain 
                   \email{p.riviere@oan.es}
                   \and Laboratoire d'Astrophysique de Bordeaux, Univ. Bordeaux, CNRS, B18N, all\'ee Geoffroy Saint-Hilaire, 33615 Pessac, France 
                   \and Univ. Grenoble Alpes, CNRS, Institut de Plan\'etologie et d'Astrophysique de Grenoble (IPAG), 38000 Grenoble, France 
                   \and Institut de Radioastronomie Millim\'etrique (IRAM), 38406 Saint-Martin d'H\`eres, France 
                   \and CNRS / Institut de Recherche en Astrophysique et Plan\'etologie, 14 avenue Edouard Belin, F-31400 Toulouse, France 
                   \and European Southern Observatory (ESO), Alonso de C\'ordova 3107, Vitacura, Casilla 19001, Santiago de Chile, Chile 
                    \and Joint ALMA Observatory, Alonso de C\'ordova 3107, Vitacura, Santiago 763-0355, Chile 
                    \and European Southern Observatory, Karl-Schwarzschild-Str 2, 85748 Garching, Germany 
                    \and Institut de Radioastronomie Millim\'etrique, 300 rue de la Piscine, F-38406 Saint-Martin d'H\`eres, France 
   }
   \authorrunning{Rivi\`ere-Marichalar et al.}
   \date{}

 \abstract 
{The sulfur abundance is poorly known in most environments. Yet, deriving the sulfur abundance is key to understanding the evolution of the chemistry from molecular clouds to planetary atmospheres. We present observations of $\rm H_2S$ $\rm 1_{10}-1_{01}$ at 168.763 GHz toward the Herbig Ae star AB Aur.}
{We aim to study the abundance of sulfuretted species toward AB Aur and to constrain how different species and phases contribute to the sulfur budget.}
{We present new NOrthern Extended Millimeter Array (NOEMA) interferometric observations of the continuum and  $\rm H_2S$ $\rm 1_{10}-1_{01}$ line at 168.763 GHz toward AB Aur. We derived radial and azimuthal profiles and used them to compare the geometrical distribution of different species in the disk. Assuming local thermodynamical equilibrium (LTE), we derived column density and abundance maps for H$\rm _2$S, and we further used \texttt{Nautilus} to produce a more detailed model of the chemical abundances at different heights over the mid-plane at a distance of  r=200 au.}
{We have resolved H$\rm _2$S emission in the AB Aur protoplanetary disk. The emission comes from a ring extending from 0.67$\arcsec$ ($\rm \sim$109 au) to 1.69$\arcsec$ ($\rm \sim$275 au). Assuming T=30 K, n$\rm _H$=10$\rm ^{9}~cm^{-3}$, and an ortho-to-para ratio of three, we derived a column density of (2.3$\rm \pm$0.5)$\rm \times 10^{13}~cm^{-2}$. Under simple assumptions, we derived an abundance of (3.1$\rm \pm$0.8)$\rm \times 10^{-10}$ with respect to H nuclei, which we compare with \texttt{Nautilus} models to deepen our understanding of the sulfur chemistry in protoplanetary disks. Chemical models indicate that H$\rm _2$S is an important sulfur carrier in the solid and gas phase. We also find an important transition at a height of $\rm \sim$12 au, where the sulfur budget moves from being dominated by ice species to being dominated by gas species.}
{We confirm that present-day models still struggle to simultaneously reproduce the observed column densities of the different sulfuretted species, and the observed abundances are still orders of magnitude away from the cosmic sulfur abundance. Studying sulfuretted species  in detail in the different phases of theinterstellar medium is key to solving the issue.} 

\keywords{Astrochemistry -- ISM: abundances  -- ISM: molecules --
   stars: formation}
\titlerunning{AB Aur, a Rosetta stone for studies of planet formation (II)}
\maketitle

\section{Introduction} 

\begin{figure*}[t!]
\begin{center}
 \includegraphics[width=0.3\textwidth,trim = 0mm 0mm 0mm 0mm,clip]{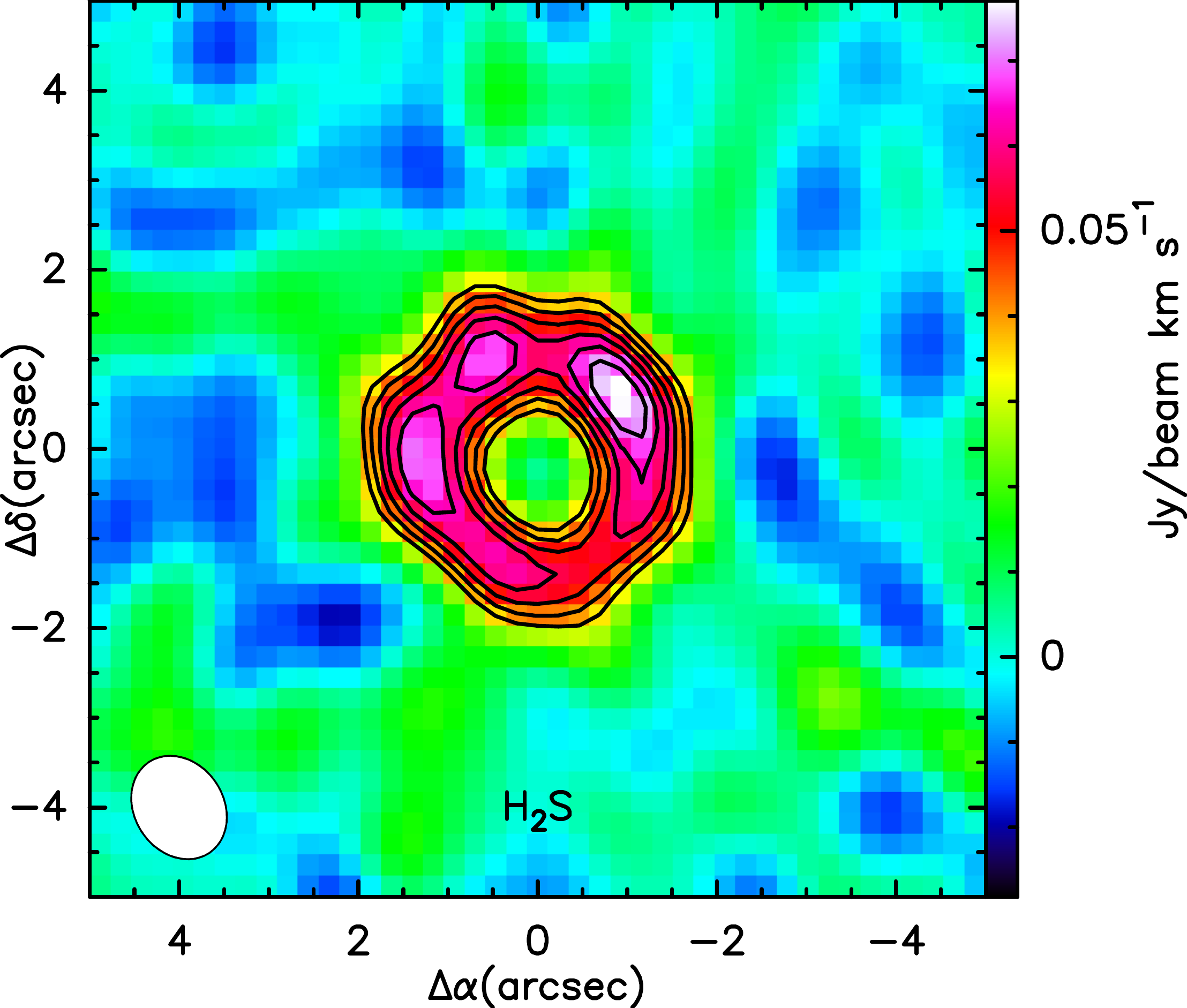}\hspace{0.3cm}\includegraphics[width=0.3\textwidth,trim = 0mm 0mm 0mm 0mm,clip]{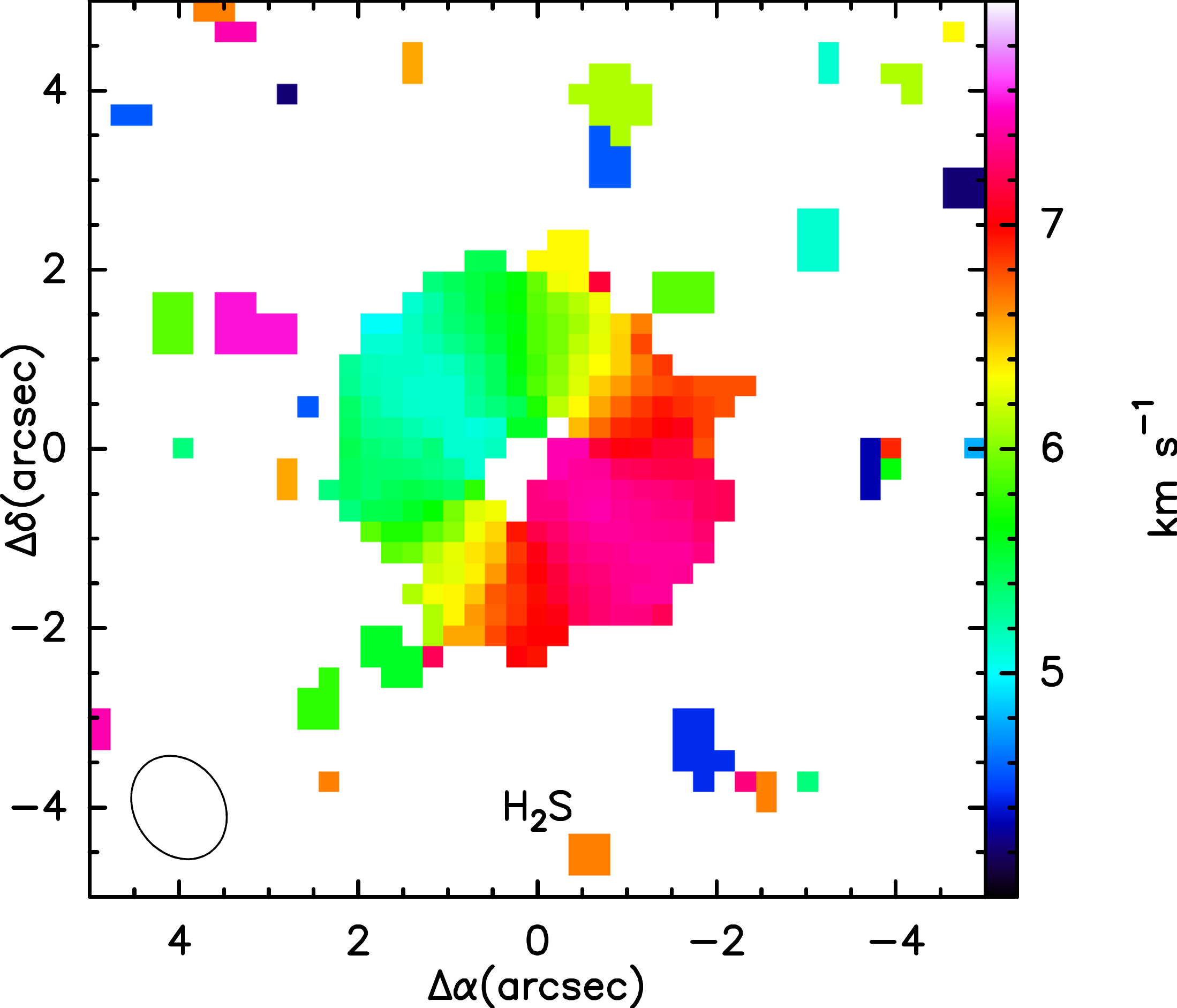}\hspace{0.3cm}\includegraphics[width=0.3\textwidth,trim = 0mm 0mm 0mm 0mm,clip]{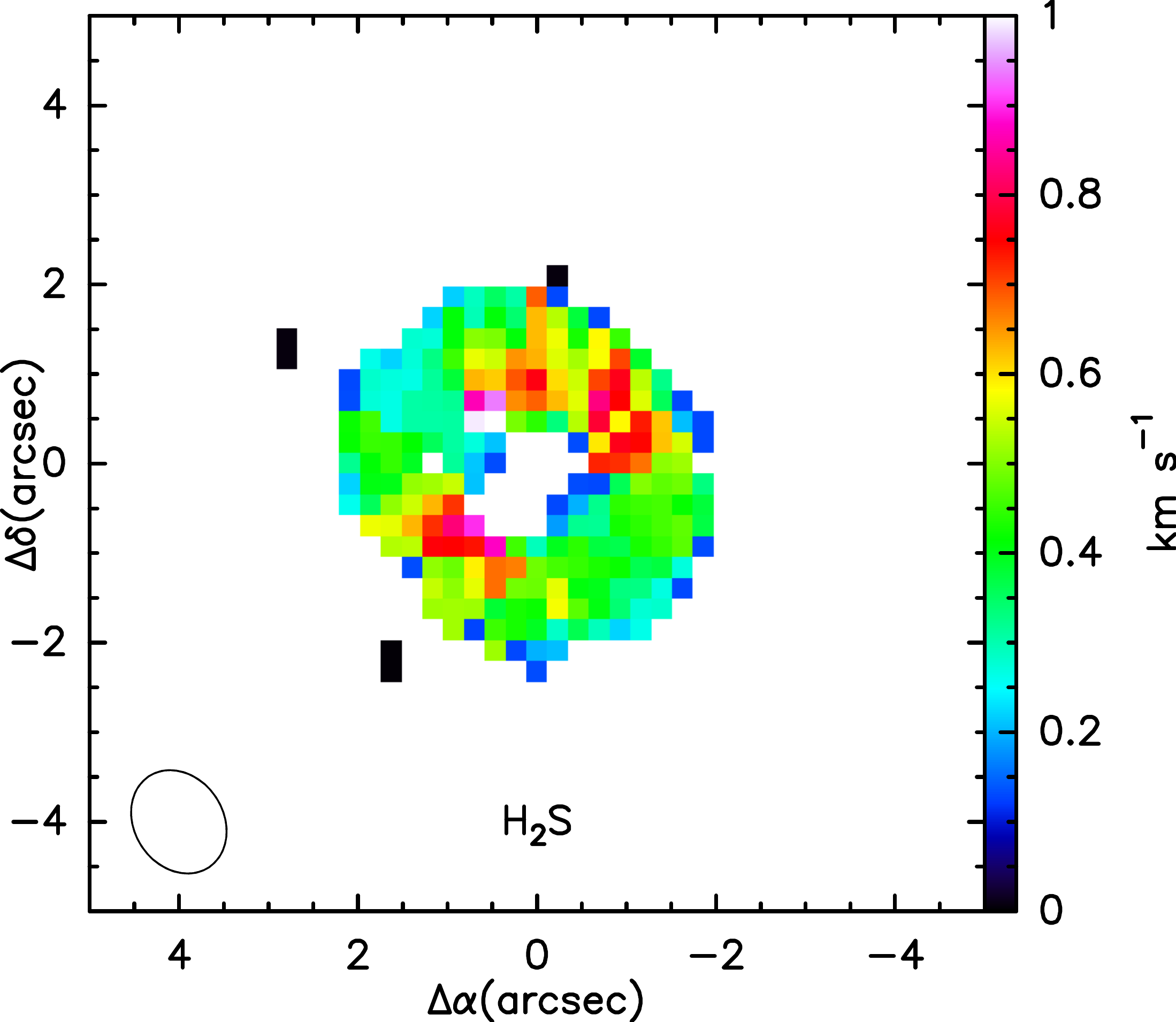}\\
  \caption{From left to right, zeroth-, first-, and second-moment maps of the H$\rm_2$S 1$\rm _{10}$-1$\rm _{01}$ emission line toward AB Aur. The synthesized beam is shown in the lower left corner of each plot.}
 \label{Fig:moment_maps}
\end{center}
\end{figure*}

\begin{figure}[t!]
\begin{center}
 \includegraphics[width=0.45\textwidth,trim = 0mm 2mm 0mm 0mm,clip]{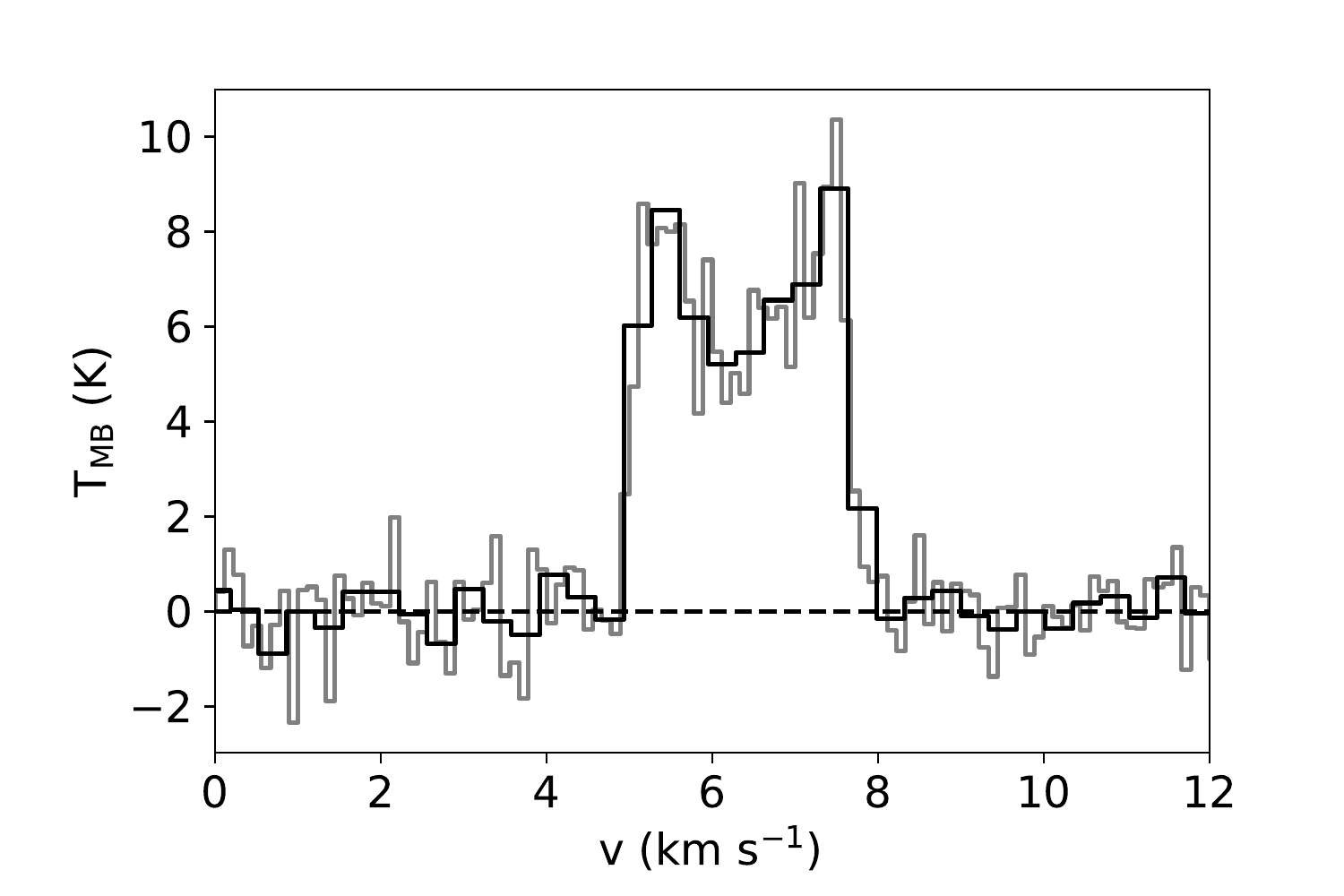}\\
  \caption{Stacked $\rm H_2S$ 1$\rm _{01}-1_{10}$spectrum of AB Aur inside the 5$\rm \sigma $ emission contour levels. The gray spectrum depicts the observed spectrum, and the black one  depicts the same after rebinning. }
 \label{Fig:H2S_spectrum}
\end{center}
\end{figure} 

Protoplanetary disks are the birthplace of planets, implying that a planet's composition is, at least partially, inherited from the disk. The protoplanetary disk chemical composition, in turn, is inherited from the natal molecular cloud, but modified by the prevailing physical conditions of protoplanetary disks. To study the chemical complexity of planetary atmospheres, one must understand that of protoplanetary disks. A topic of major interest due to its possible implications for the emergence of life is the chemical evolution of molecules that carry the six key elements in organic chemistry: H, C, O, N, S, and P.

Sulfur is the tenth most abundant element in the Universe and plays a crucial role in biological systems. The abundance toward the Sun \citep[S/H$\sim$1.5$\times$10$^{-5}$]{Asplund2009} agrees well with the value derived toward Orion B stars for example \citep[S/H$\sim$1.4$\times$10$^{-5}$,][]{Daflon2009}. However, its chemistry is poorly understood in interstellar environments, and sulfuretted molecules are not as abundant as expected in the interstellar medium (ISM). While sulfur abundance is close to the cosmic value in the diffuse ISM and photon-dominated regions (PDRs)  \citep{Goicoechea2006,Howk2006,Goicoechea2021b}, it is strongly depleted in the dense molecular gas ($\rm >$10$\rm ^{4}~cm^{-3}$), where only 0.1\% of the sulfur cosmic abundance is observed \citep{Tieftrunk1994, Wakelam2004, Vastel2018}. A study of the sulfur content in dense cores by \cite{Hily-Blant2022} provides evidence for a progressive depletion of sulfur with age, as well as with density for a given core. It is expected that most of the sulfur is then locked on the icy mantles of dust grains in dense cores \citep{Millar1990, Ruffle1999, Vidal2017, Laas2019}. However, only solid OCS has been unambiguously detected on the ice mantles of these objects thanks to the strength of its infrared band \citep{Geballe1985,Palumbo1995}. Furthermore, SO$\rm _2$ has also been tentatively detected \citep{Boogert1997}. Since only upper limits for the solid H$_2$S abundance could be derived thus far \citep{JimenezEscobar2011} and since detection is hampered by the strong overlap between the 2558 cm$^{-1}$ band and the methanol bands at 2530 and 2610 cm$^{-1}$, the main gas and solid phase sulfur reservoirs thus remain unknown.

Sulfur-bearing species have been widely detected in the Solar System. In comets, they are detected mostly in the form of  H$_2$S and S$_2$ \citep{Mumma2011}. Hale Bopp showed a large variety of detections, including CS and SO \citep{Boissier2007}. Furthermore, CS was also detected toward the comets C/2012 F6 (Lemmon) and C/2014 Q2 (Lovejoy) \citep{Biver2016}. Using the Rosetta Orbiter Spectrometer for Ion and Neutral Analysis (ROSINA \citep{Balsiger2007}) on board Rosetta, H$_2$S, S, SO$_2$, SO, OCS, H$_2$CS, CS$_2$, and S$_2$ were detected in the coma \citep{LeRoy2015} of 67P/Churyumov-Gerasimenko. In addition, S$_3$, S$_4$, CH$_3$SH, and C$_2$H$_6$S were also detected \citep{Calmonte2016}. Considering the variety of S species detected, the mean  abundance of H$_2$S relative to \ce{H2O} remains around 1.5\% \citep{Bockelee2017} with values ranging from $\rm 10^{-3}$ to 0.1.

Protoplanetary disks (PPDs) are expected to strongly affect the abundance of different sulfuretted species given the multiple chemical reprocessing mechanisms at work in them. Therefore, they are key to understanding the huge differences (at a factor of a thousand) in the sulfur abundances between the diffuse ISM and dense molecular gas. Searches for S-bearing molecules in PPDs have provided few detections. Thus far, only one S species, CS, has been widely detected \citep{Dutrey2011, Guilloteau2012, Pacheco2015, Guilloteau2016, Podio2020, Podio2020b, Rosotti2021, Riviere2021, Nomura2021}.  We note that H$_2$CS was detected in MWC 480 \citep{LeGal2019}, and H$_2$S was detected in GG Tau  \citep{Phuong2018} and in four protoplanetary disks in Taurus \citep{Riviere2021}. Furthermore, SO has been detected in a few PPDs \citep{Pacheco2016,Booth2018,Booth2021,LeGal2021}, and SO$\rm _2$ has been observed toward IRS 48 as well \citep{Booth2021}. Recently, CCS was detected toward GG Tau \citep{Phuong2021}, but the low signal-to-noise ratio (s/n) precluded the analysis of the spatial distribution. The authors could not reproduce the observed CCS column density together with other sulfuretted species, supporting the idea that astrochemical sulfur networks are incomplete. Similar problems have been encountered in CS studies, such as in the study of cold cores by \cite{Navarro2020}, where models overpredicted CS abundances by a factor of 10. Sulfur astrochemical networks have improved over the last years \citep{Fuente2016, Fuente2017S2H, LeGal2017, Vidal2017, Fuente2019, Laas2019}, including new formation and destruction routes, but work is still needed to solve the issue. 

\begin{figure*}[t!]
\begin{center}
\begin{minipage}[c]{0.8\textwidth}
 \includegraphics[width=1.0\textwidth]{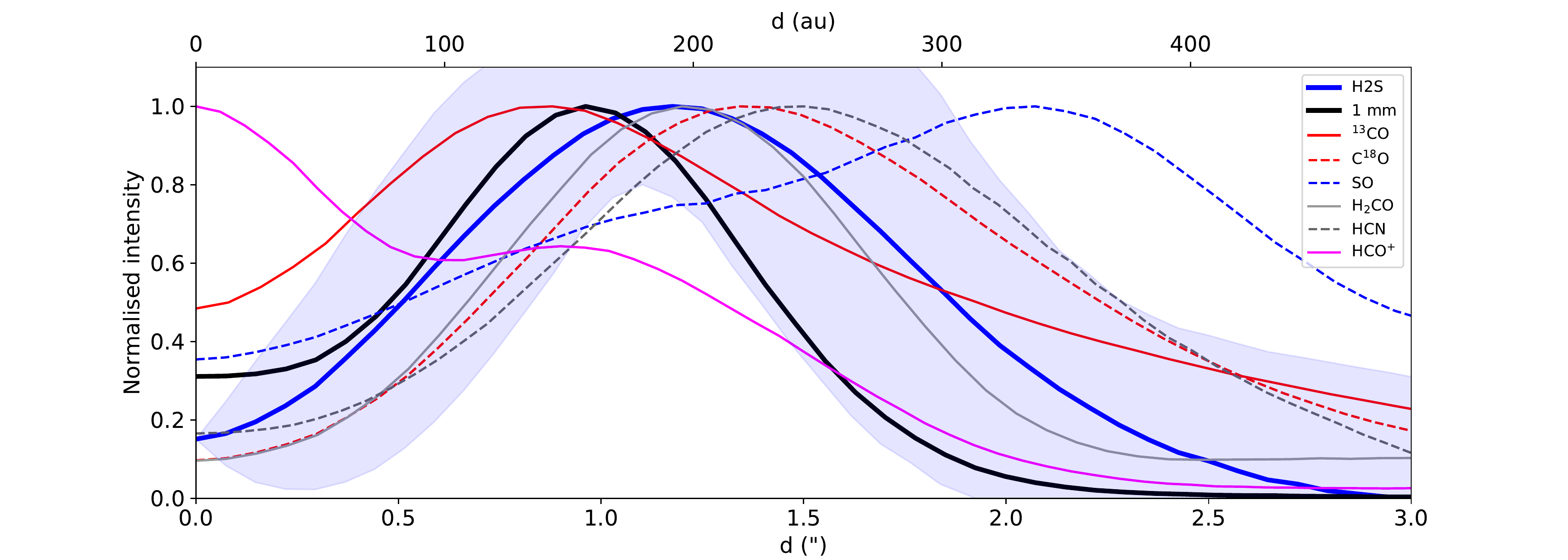}\\
\end{minipage}\hfill
\begin{minipage}[c]{0.2\textwidth} 
  \caption{H$\rm_2$S azimuthally averaged radial profile compared to previous observations by this team. All the intensity maps have been convolved to match the beam size of our H$\rm_2$S emission line maps. The light-blue shaded area depicts the \ce{H2S} radial profile plus or minus uncertainties.}
\end{minipage}
 \label{Fig:radial_profile}
\end{center}
\end{figure*}

The Herbig star AB Aur is a well-known Herbig A0-A1 \citep{Hernandez2004} star that hosts a transitional disk. The system is located at 162.9 pc from the Sun \citep{GAIA2018}, and it is perfectly suited to study the spatial distribution of gas and dust in circumstellar environments in detail. The disk extends out to a radius of 2.3$\arcsec$ ($\rm \sim$373 au) in the continuum \citep[][]{Riviere2020}. Molecular emission from species such as CS, C$\rm ^{18}$O, SO, \ce{H2CO}, and \ce{H2S} is observed with radii of emission peaks ranging from  0.9$\arcsec$ to 1.4$\arcsec$. The disk depicts a series of features that could be linked to planet formation, such as prominent spiral arms at the near-IR and radio wavelengths. The system also presents a cavity in continuum emission \citep{Pietu2005,Tang2012,Fuente2017} that extends to $\sim$70-100 au. \cite{Tang2012} discovered a compact source inside the cavity, which is most likely an inner disk well suited to explain the strong accretion observed toward the source \citep{GarciaLopez2006, Salyk2013}. \cite{Rodriguez2014} detected a radio jet consistent with the high levels of accretion observed. Inside the dust cavity, prominent CO spiral arms were observed by \cite{Tang2017} using the Atacama Large Millimeter/submillimeter Array (ALMA). Other species have been detected toward AB Aur by different studies, including CO, SO, \ce{HCO+}, HCN, and \ce{H2CO}  \citep{Schreyer2008, Pietu2005, Tang2012, Tang2017, Fuente2010, Pacheco2015, Pacheco2016, Riviere2019, Riviere2020}.  The first detection of SO in a protoplanetary disk was reported, in fact, also toward AB Aur \citep{Fuente2010, Pacheco2015, Pacheco2016}. In \cite{Riviere2019}, we presented high angular resolution the NOrthern Extended Millimeter Array (NOEMA) observations of HCN and \ce{HCO+}, with a beam size of 0.4$\arcsec$. The \ce{HCO+} map depicts an outer disk with decay in intensity coincident with the dust cavity,  a compact source toward the center, and a bridge of material that connects the outer disk with the compact source in the center. In paper I of this series, we presented the results of a NOEMA spectral survey in AB Aur \citep[Paper I, ][]{Riviere2020}, where we were able to obtain zeroth-, first-, and second-moment maps, opacity maps, temperature maps, and column density maps of the transitions and species surveyed. These species included $\rm ^{12}CO$, $\rm ^{13}CO$, C$\rm ^{18}O$, \ce{H2CO}, and SO. We derived a mean disk temperature of 39 K, and column densities in the range 10$\rm ^{12}$ to 5$\times$10$\rm ^{13}~cm^{-2}$ for H$\rm _2$CO, SO, HCO$^+$ and HCN, and $\rm \sim 10^{17}~cm^{-2}$ for $\rm ^{13}CO$. We computed a gas-to-dust mass ratio map of AB Aur and showed it to range from 10 to 40 across the disk, with larger values close to the disk's inner edge. Such values are between two times and one order of magnitude smaller than the typical value of 100 found in the ISM. The minimum in the gas-to-dust mass ratio was coincident with the peak of the continuum emission, indicating a particularly gas-poor dust trap. We produced radially and azimuthally averaged profiles of line intensity, temperature, and column densities. Such profiles demonstrate the strong chemical segregation observed in the source, with differences as high as 100 au in the position of their peaks.

In this paper, we present the first detection of \ce{H2S} toward AB Aur obtained with NOEMA. In Sect. \ref{Sect:obs_data_red} we introduce the observations setup and summarize the reduction process. In Sect. \ref{Sect:results} we present our results, including estimates of the \ce{H2S} column density and abundance. In Sect. \ref{Sect:discussion} we discuss the implications of our results for the topic of sulfur chemistry in protoplanetary disks. Finally, in Sect. \ref{Sect:summary} we summarize our conclusions. 

\section{Observations and data reduction}\label{Sect:obs_data_red}
We observed the H$\rm _2$S 1$\rm _{10}$-1$\rm _{01}$ transition at 168.763 GHz toward AB Aur with NOEMA between May and July 2020 using EMIR receivers \citep{Carter2012}. The source was observed for eight hours in configuration 8, with baselines ranging from 18 to 315 m. The data reduction and map synthesis was performed using \texttt{GILDAS}\footnote{See \texttt{http://www.iram.fr/IRAMFR/GILDAS} for  more information about the GILDAS  software.}\texttt{/MAPPING}. We used the PolyFix correlator centered at 161.5 GHz with a bandwidth of 8 GHz per sideband. A chunk with a spectral resolution of 62.5 kHz was placed at 168.763 GHz to observe H$\rm _2$S 1$\rm _{10}$-1$\rm _{01}$, reaching a velocity resolution of 0.1 km s$\rm ^{-1}$. The final map was built using robust weight, allowing us to reach a beam size of 1.19$\arcsec \times$0.97$\arcsec$ (193 au $\rm \times$ 158 au), with PA=34$\rm ^{\circ}$ the PA defined as positive from N to E.

\begin{figure}[t!]
\begin{center}
 \includegraphics[width=0.5\textwidth,trim = 0mm 0mm 0mm 0mm,clip]{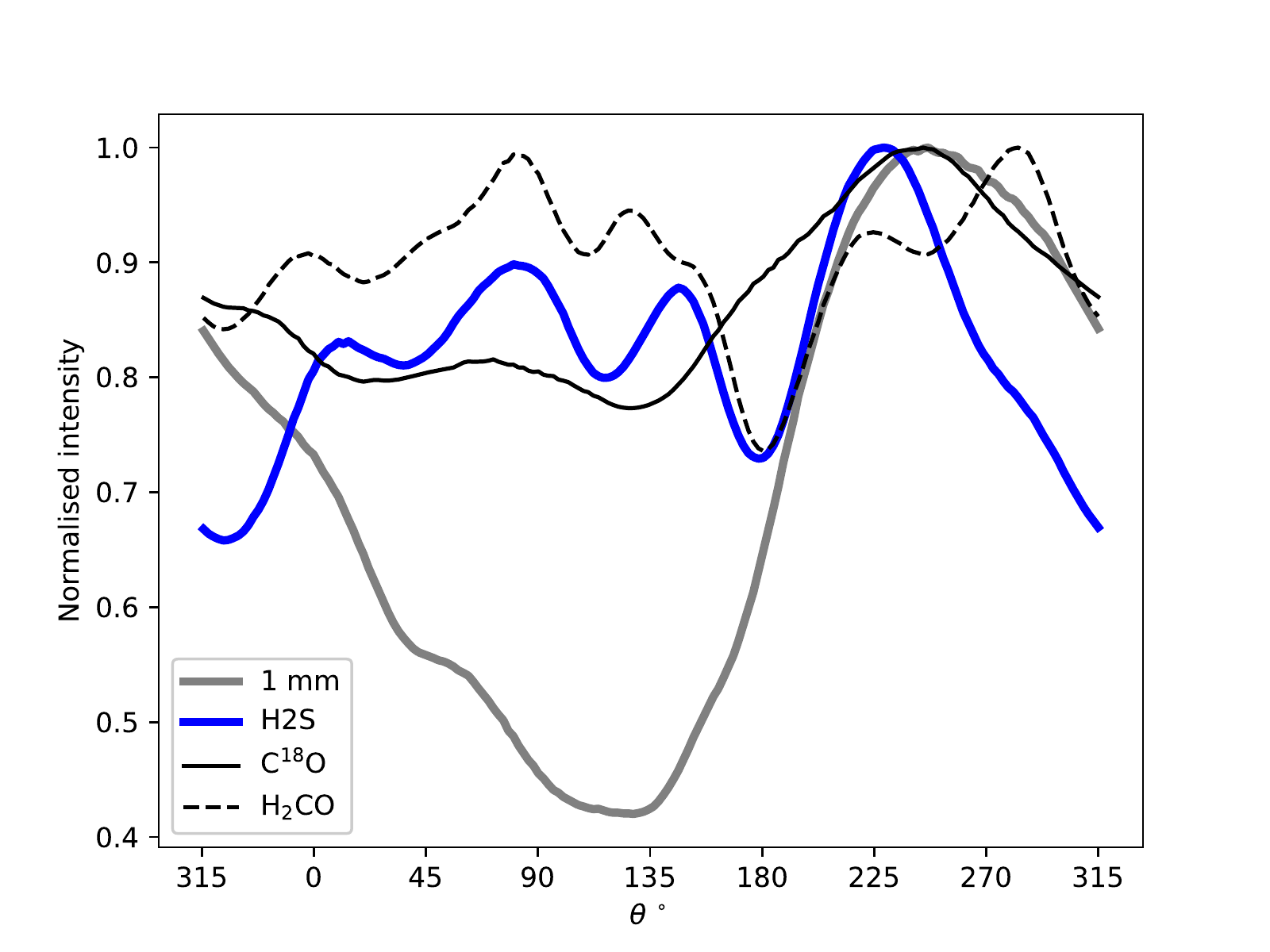}\\
  \caption{H$\rm_2$S azimuthal cuts compared to previous observations by this team. All the intensity maps have been convolved to match the beam size of our H$\rm_2$S emission line maps. The azimuthal profiles shown were computed at the distance of the emission peak of each species. The profiles were normalized with respect to their respective peaks.}
 \label{Fig:azimuthal_profile}
\end{center}
\end{figure}

\section{Results}\label{Sect:results}

We show in Fig. \ref{Fig:moment_maps} the resulting zeroth-, first- and second-moment maps of H$\rm _2$S 1$\rm _{10}$-1$\rm _{01}$ at 168.763 GHz. The integrated intensity map depicts a ring of H$\rm _2$S emission extending from  $\rm \sim$0.7$\arcsec$ ($\rm \sim$114 au) to $\rm \sim$1.7$\arcsec$ ($\rm \sim$277 au). The ring shows strong azimuthal asymmetries, with a peak that is roughly coincident with the position of the continuum peak \citep{Fuente2017} and the position of the H$\rm _2$CO \citep{Riviere2020} and HCN \citep{Riviere2019} integrated emission peaks. 

In Fig. \ref{Fig:H2S_spectrum} we show the stacked spectrum inside the 5$\sigma$ emitting contours. The spectrum depicts the two-peaked profile characteristic of a rotating disk. The two peaks show similar intensities (8.3 K and 8.8 K), and line widths (0.4 and 0.54 km s$\rm ^{-1}$). The blue-shifted component peaks at 5.4 km s$\rm ^{-1}$, and the red-shifted one at 7.0 km s$\rm ^{-1}$. 

In Fig. \ref{Fig:radial_profile} we show the azimuthally averaged radial profile of the H$\rm _2$S 1$\rm _{10}$-1$\rm _{01}$ emission line and compare it to the radial profile of other transitions that have been observed with high spatial resolution toward AB Aur, including $\rm ^{12}$CO, $\rm ^{13}$CO, C$\rm ^{18}$O, SO, and H$\rm _2$CO \citep{Riviere2020}, and HCO$\rm ^+$ and HCN \citep{Riviere2019}. These maps have been deprojected, assuming an inclination angle  i = 26$\rm ^{\circ}$ and position angle PA = -37$\rm ^{\circ}$. The maps were also convolved with the H$\rm _2$S beam for a better comparison of the radial profiles. The \ce{H2S} emission ring shows a peak at $\rm \sim$1.2$\arcsec$ ($\rm \sim$195 au), with emission in excess over the RMS extending from $\rm \sim$0.51$\rm \arcsec$ ($\rm \sim$83 au) to $\rm \sim$2.35$\rm \arcsec$ ($\rm \sim$383 au). The average disk width is 1.84$\rm \arcsec$ ($\rm \sim$300 au). \ce{H2S} shows a radial profile that matches that of \ce{H2CO}. In a recent study of \ce{H2S} emission in Taurus disks, a tentative correlation between \ce{H2S} and \ce{H2CO} was shown \citep{Riviere2021}; our maps now point to a very similar spatial origin for both emission lines, increasing the observational support for such a correlation. Furthermore, both species have formation routes on the surface of grains. In Fig. \ref{Fig:azimuthal_profile} we show cuts in azimuth of the \ce{H2S}, \ce{H2CO}, C$\rm ^{18}$O, and continuum emission maps at the radius of their respective emission peaks. The similarity between \ce{H2S} and \ce{H2CO} is again prominent, with local maxima and minima at almost the same azimuths. The azimuthal contrast ratio of \ce{H2S} at the distance to the emission peak is 1.5 $\rm \pm$ 0.3.

\begin{figure}[t!]
\begin{center}
 \includegraphics[width=0.45\textwidth,trim = 10mm 0mm 0mm 0mm,clip]{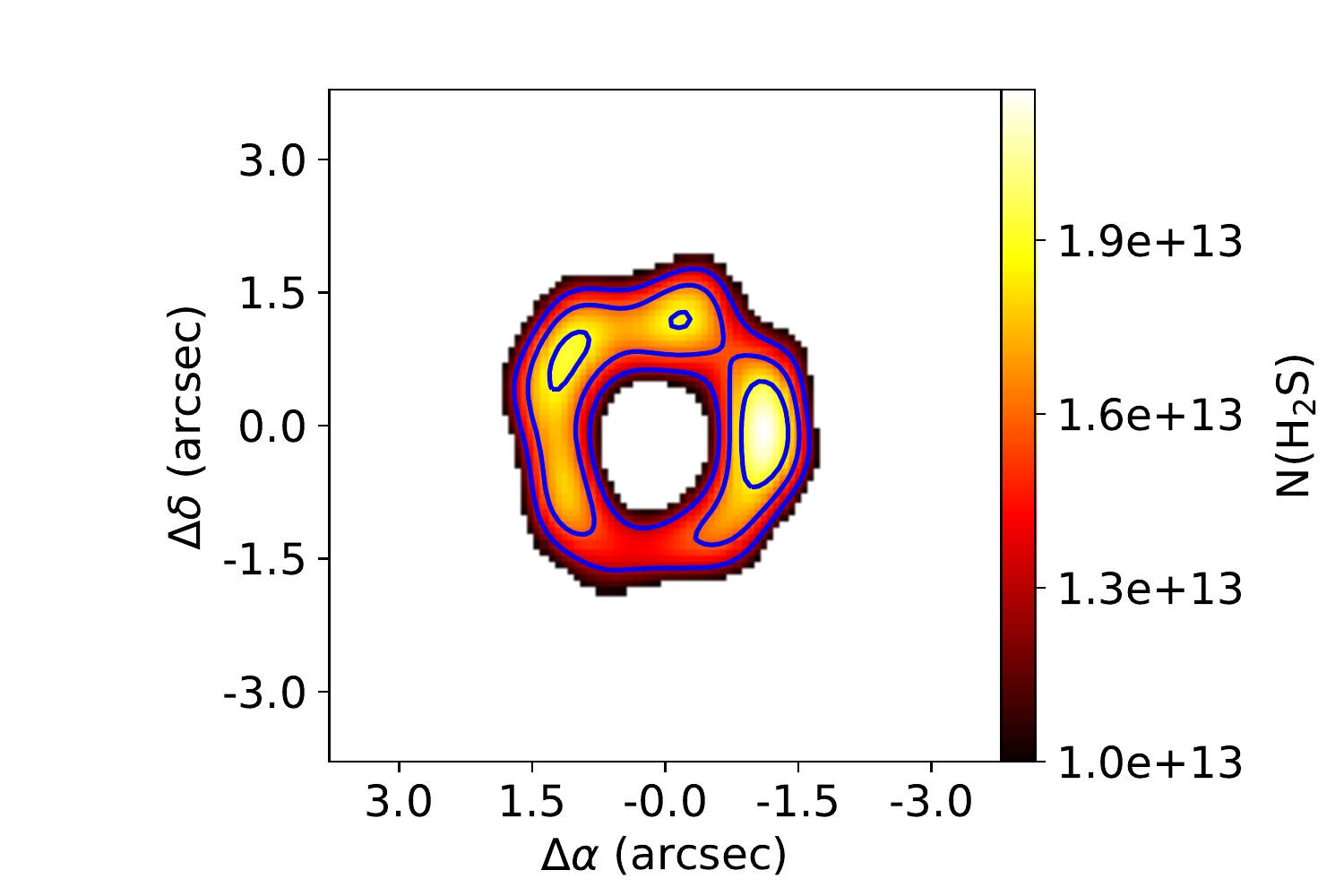}\\
  \caption{\ce{H2S} column density map computed assuming LTE and T$\rm _k$ = 30 K and ortho-to-para ratio of 3. The map has been deprojected assuming an inclination angle i = 26$\rm ^{\circ}$ and position angle PA = -37$\rm ^{\circ}$. }
 \label{Fig:N_H2S_map}
\end{center}
\end{figure}

Assuming local thermodynamical equilibrium (LTE) we compute an \ce{H2S} column density map. Since we have observed only one o-\ce{H2S} transition ($\rm 1_{10}-1_{01}$) no rotational diagram could be computed and we thus assumed a fixed disk temperature. For consistency, we adopted a temperature of 30 K as in our previous work \citep{Riviere2020}, allowing the comparison between \ce{H2S} and the species surveyed in this work. The mean \ce{o-H2S} column density is (1.5$\rm \pm$0.3)$\rm \times 10^{13}~cm^{-2}$, with a minimum value of 1.0$\rm \times 10^{13}~cm^{-2}$ and a maximum of 2.2$\rm \times 10^{13}~cm^{-2}$. Assuming that \ce{H2S} behaves like \ce{H2O}, we assign an ortho-to-para ratio of 3 \citep{Hama2016}, and derive a \ce{H2S} mean column density of (1.9$\rm \pm$0.4)$\rm \times 10^{13}~cm^{-2}$. The resulting deconvolved column density map is shown in Fig. \ref{Fig:N_H2S_map}. The peak in column density is coincident with the dust trap. There are other two local maxima at PA$\rm \sim 90 ^{\circ}$ and $\rm \sim 135 ^{\circ}$.

\section{Modeling the sulfur budget in AB Aur}\label{sect:mod_sulfur_budget}
To gain further insight into the AB Aur sulfur budget, we also computed a \texttt{Nautilus} 1D model. \texttt{Nautilus} computes the evolution of molecular abundances for a set of initial abundances and physical parameters using gas-phase, gas-grain, and surface chemistry reactions \citep{Semenov2010, Loison2014, Wakelam2014, Reboussin2015}. The parameters include gas and dust temperature, gas density, cosmic ray molecular hydrogen ionization rate $\rm \zeta_{H_{2}}$, and extinction, among others. A refinement of the code considers grain surface and grain mantle reactions by using a three-phase structure that includes gas, grain surfaces, and grain mantles \citep{Ruaud2016,Wakelam2017}. For a more detailed description of the code see \citet{Ruaud2016} and \citet{Wakelam2017}.

\begin{table}[t]
\caption{Nautilus model input parameters for the molecular cloud prephase and the protoplanetary disk at r=200 au.}
\label{tab:model_parameters}
\begin{center}
\begin{tabular}{ll}
\hline \hline
Parameter & Value \\
\hline
\multicolumn{2}{c}{Molecular cloud}\\
\hline
$\rm T_{gas}$ (K) & 10 \\
$\rm T_{dust}$ (K) & 10 \\
n$\rm _H$& 10$\rm ^4$\\
$\rm A_v$ & 20 \\
f$\rm _{UV}$ (Draine units) & 1 \\
$\rm \zeta_{H_{2}}$ (s$\rm ^{-1}$)& $\rm 10^{-17}$ \\
Gas-to-dust mass ratio & 100 \\
\hline
\multicolumn{2}{c}{AB Aur at r=200 au}\\
\hline
$\rm T_{mid}$ (K) & 39 \\
$\rm T_{atm}$ (K) & 65  \\
$\rm A_v$ & 2 \\
f$\rm _{UV} (Draine units) $ & 1.2$\rm \times 10^{4}$ \\
$\rm \zeta_{H_{2}}$ (s$\rm ^{-1}$)& $\rm 10^{-17}$ \\
Gas-to-dust mass ratio & 40 \\
\hline
\end{tabular}
\end{center}
\label{default}
\end{table}%

\begin{table}
\centering
\caption{Initial abundances}
\label{Table:elemental_ab}
\begin{tabular}{lcc}
\hline\hline
Species & $n_i/n_{\text{H}}$    & Reference\\
\hline
H$_2$   & 0.5 & -\\
He      & 9.0$\times$10$^{-2}$              & 1\\
C$^+$   & 1.7$\times$10$^{-4}$              & 2 \\
N       & 6.2$\times$10$^{-5}$              & 2\\  
O       & 2.4$\times$10$^{-4}$              & 3\\ 
S$^+$   & 1.5$\times$10$^{-5}$   & 4 \\
Si$^+$  & 8.0$\times$10$^{-9}$              & 4 \\
Fe$^+$  & 3.0$\times$10$^{-9}$              & 4 \\
Na$^+$  & 2.0$\times$10$^{-9}$              & 4 \\
Mg$^+$  & 7.0$\times$10$^{-9}$              & 4 \\
P$^+$   & 2.0$\times$10$^{-10}$             & 4 \\
Cl$^+$  & 1.0$\times$10$^{-9}$              & 4 \\
F$^+$   & 6.7$\times$10$^{-9}$             & 5 \\
\hline
\label{tab:elemental_ab}
\end{tabular}
\noindent
\tablefoot{
(1) \cite{Wakelam2008}; (2) \cite{Jenkins2009};
(3) \cite{Hincelin2011};
(4) \cite{Graedel1982};
(5) \cite{Neufeld2015}}

\end{table}

\begin{figure*}[h!]
\begin{center}
 \includegraphics[width=0.45\textwidth,trim = 0mm 0mm 0mm 0mm,clip]{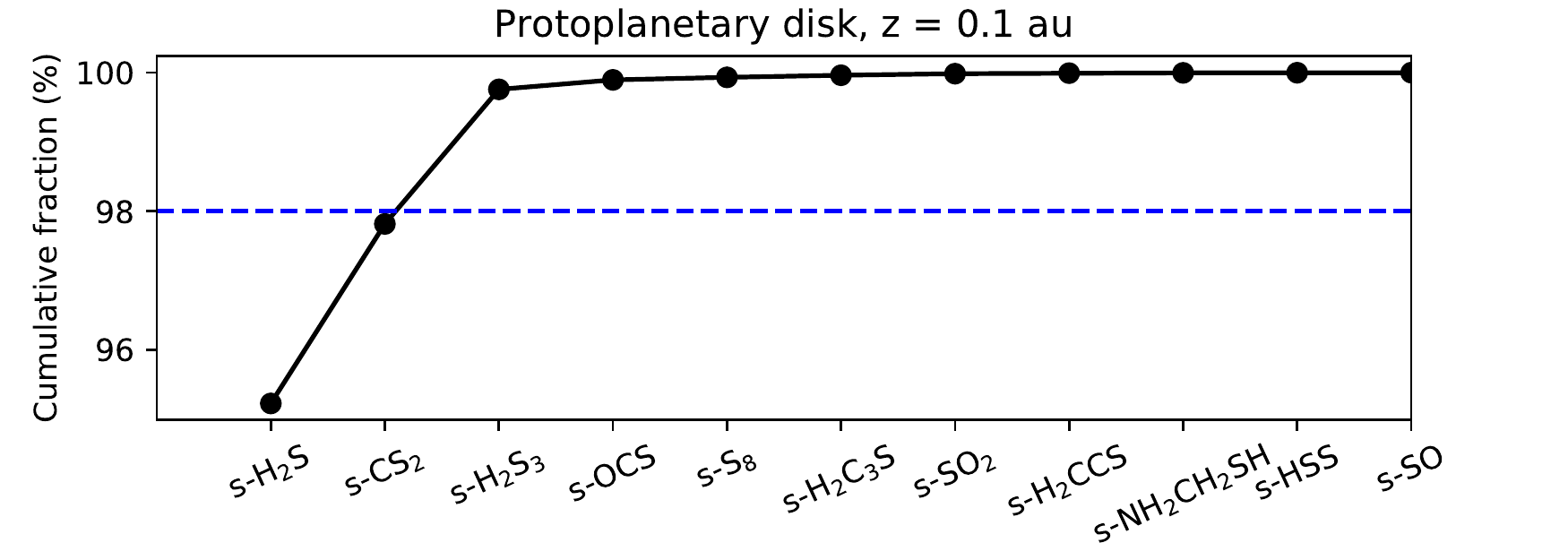} \includegraphics[width=0.45\textwidth,trim = 0mm 0mm 0mm 0mm,clip]{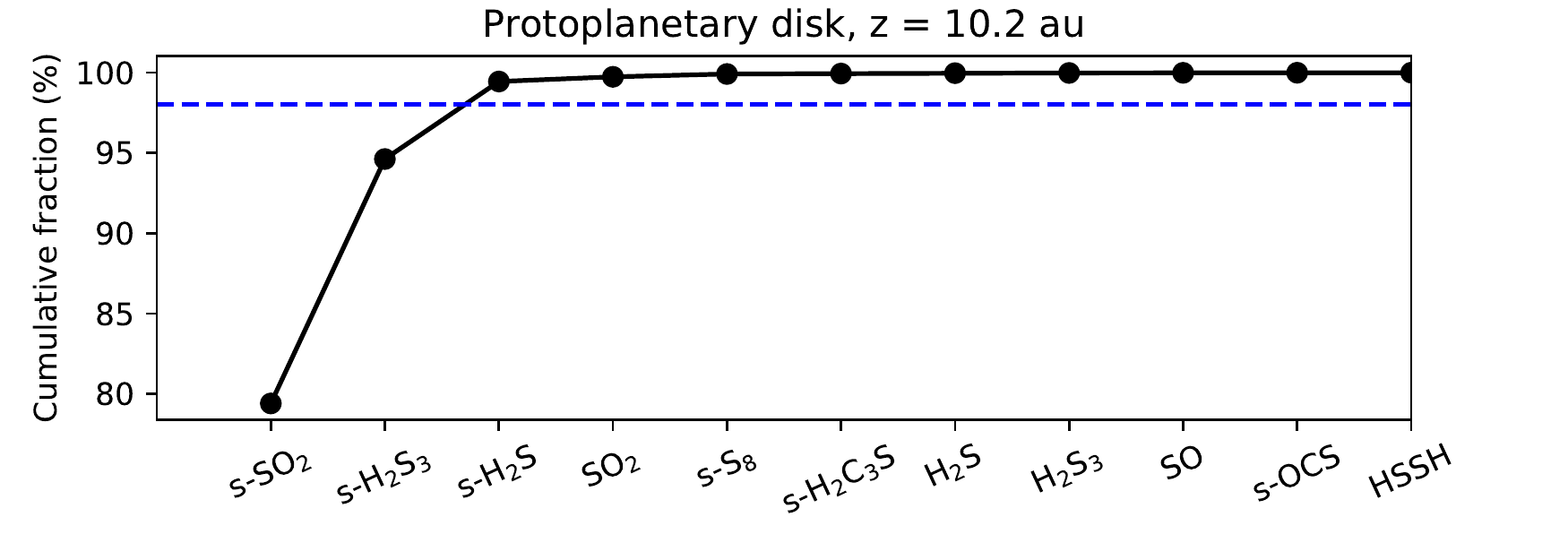} \\
 \includegraphics[width=0.45\textwidth,trim = 0mm 0mm 0mm 0mm,clip]{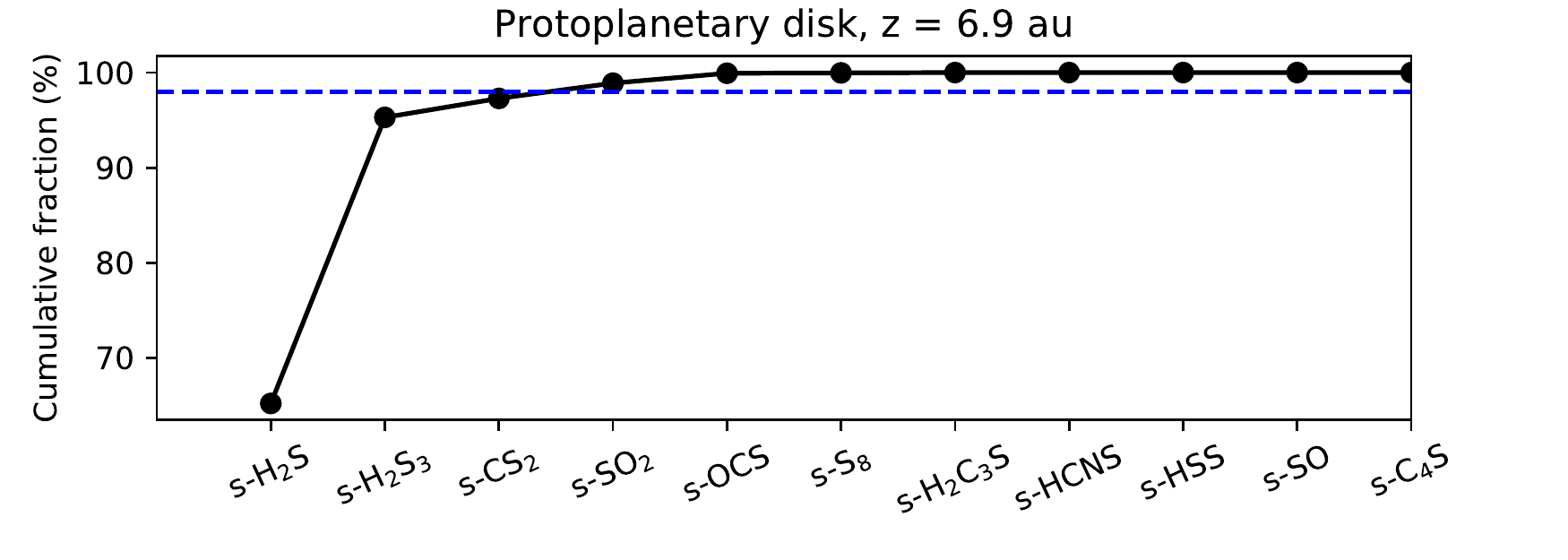} \includegraphics[width=0.45\textwidth,trim = 0mm 0mm 0mm 0mm,clip]{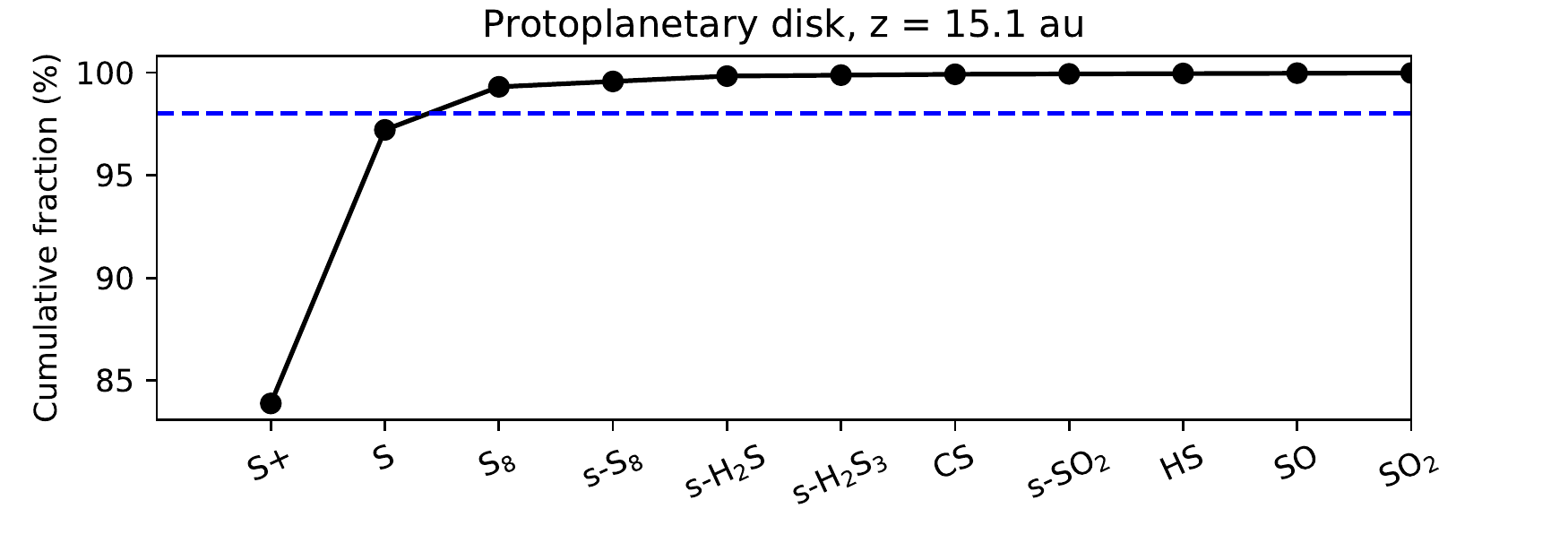}\\
 \includegraphics[width=0.45\textwidth,trim = 0mm 0mm 0mm 0mm,clip]{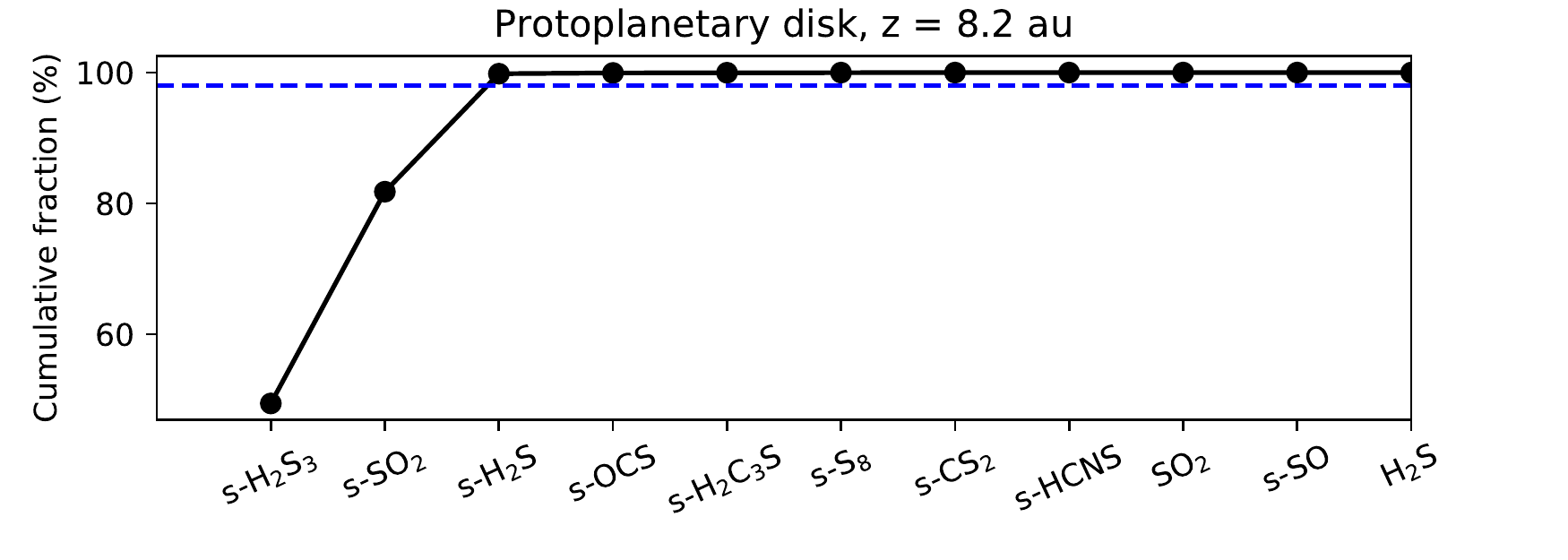} \includegraphics[width=0.45\textwidth,trim = 0mm 0mm 0mm 0mm,clip]{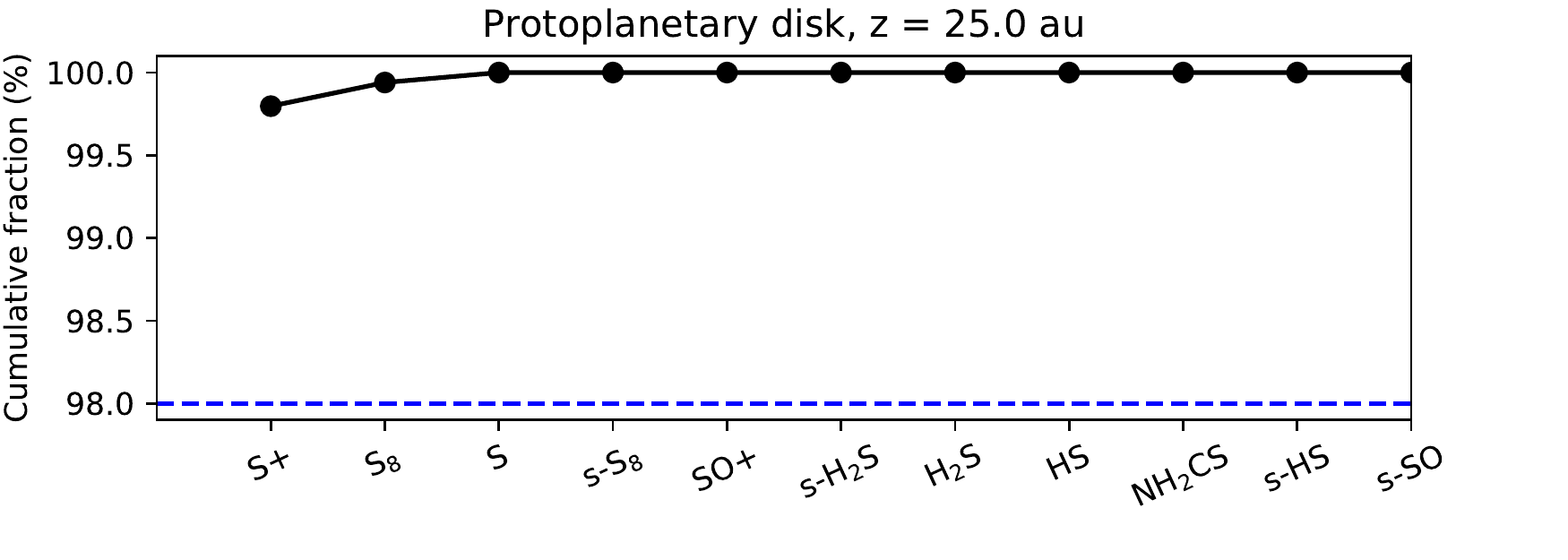} \\
  \caption{Cumulative abundance fraction of sulfuretted species from \texttt{Nautilus} models at different heights above the midplane at  r=200 au. The horizontal blue dashed line depicts the position of the 98\% cumulative fraction.}
 \label{Fig:sulfur_budget_per_height}
\end{center}
\end{figure*}

\subsection{Model setup}
The physical structure used in our models is the one used in \cite{Riviere2020}, which in turn was adapted from \cite{LeGal2019}. Given the moderate resolution of the observations, we decided not to consider the radial structure and instead focus on the vertical profile at a radius of r=200 au, representative of the \ce{H2S} ring described in Sect. \ref{Sect:obs_data_red}. Our models included both photodesorption (with a yield of 10$\rm ^{-4}$molecules photon$\rm ^{-1}$) and chemical desorption, assuming the conditions listed in Table \ref{tab:model_parameters}. Chemical desorption was included following the prescription by \cite{Garrod2007} for ice-coated grains implemented in \texttt{Nautilus}. To mimic the molecular cloud origin of the system, instead of assuming initial abundances following cosmic values, we computed a 10 K molecular cloud model and used its output abundances as the initial abundances for our 1D protoplanetary disk model. Each phase was let to evolve for 1 Myr. The input physical parameters for this model are listed in Table \ref{tab:model_parameters}, and the initial abundances used for the molecular cloud prephase are listed in Table \ref{Table:elemental_ab}. We note that we use a nondepleted value for the sulfur abundance. \texttt{Nautilus} assumes a single size of 0.1 $\rm \mu$m for the dust grains. Following \cite{Riviere2020}, we assumed a gas-to-dust mass ratio of 40 for the AB Aur protoplanetary disk. The chemical network used is the updated version of the kida.uva.2014 \citep{Wakelam2015} presented in \cite{Navarro2020} which includes an enhanced chemical network for sulfur chemistry \citep{Vidal2017}. The grid consisted of 100 heights over the mid-plane with logarithmic spacing. We note that \texttt{Nautilus} is not properly suited to model the PDR-like conditions of the uppermost layers of the disk model. Diffusion between neighboring cells was not included. Since the distinction between mantle species and surface species is mostly a computational parametrization in \texttt{Nautilus}, in the following we consider grain surfaces and mantles as a single phase, called grain surface and preceded by the prefix g-, which is computed by adding up abundances in each of the solid phases (mantle and surface of grains) for each species.

For the sake of clarity, we summarize in the following the main equations describing our model. The vertical temperature profile at r=200 au follows the prescription by \cite{Dartois2003} used in \cite{Rosenfeld2013}, described by the following equation:

\begin{equation}
\small
T(z) = \left\{
    \begin{array}{ll}
        T_{\rm{mid}}+(T_{\rm{atm}}-T_{\rm{mid}})\left[ \sin \left(\frac{\pi z}{2z_q}\right) \right]^{2\delta}&\mbox{if} \, z<z_q\\
        T_{\rm{atm}}&\mbox{if} \, z\ge z_q,
    \end{array}
\right.
\label{eq:T_z}
\end{equation}
where $\rm T_{atm}$ and $\rm T_{mid}$ are the temperature of the disk atmosphere and the disk mid plane at r= 200 au, respectively.

\begin{eqnarray}
T_{\rm{mid}}=T_{\mathrm{mid},R_c}\,\left(\frac{{\rm 200~au}}{R_{\rm{c}}}\right)^{-q},\\
T_{\rm{atm}}=T_{\mathrm{atm},R_c}\,\left(\frac{{\rm 200~au}}{R_{\rm{c}}}\right)^{-q}
,\end{eqnarray}
where $R_c$ is a characteristic radius \citep[$R_c$=98 au,][]{Riviere2020}, and $z_q=4H$, where H is the pressure scale height.

\begin{equation}
H=\sqrt{\frac{k_{\rm B} \, T_{\rm mid} \,r^3}{\mu \,m_{\rm H}\, G \,M_{\star}}},
\end{equation}
where $k_{\rm{B}}$ is the Boltzmann constant, $\mu=2.4$ is the mean molecular weight of the gas, $m_{\rm{H}}$ is the proton mass, and $M_\star$ is the mass of the central star. The mid-plane temperature $T_{\rm mid}$ is described by equation

\begin{equation}
    T_{\rm{mid}}(r)\approx \left(\frac{\varphi L_\star}{8\pi r^2 \sigma_{\rm{SB}}}\right)^{1/4},
    \label{eq:Tmid_Rc}
\end{equation}
where $L_\star$ is the stellar luminosity, $\varphi$ is the flaring index, and $\sigma_{\rm{SB}}$ is the Stefan-Boltzman constant. We assumed that the dust has the same temperature as the gas. The vertical profile of the density is described by

\begin{equation}
\rho (z) =\rho_0 e^{-\frac{z^2}{2H^{2}}}
,\end{equation}
where $\rho_0$ is the mid-plane density of the gas at 200 au. Finally, the UV flux at a given radius is given by equation

\begin{equation}
f_{\rm{UV}}=\frac{0.5 f_{\rm{UV},R_c}}{\left(\frac{r}{R_c}\right)^2+\left(\frac{4\rm{H}}{R_c}\right)^2}.    
\end{equation}

\begin{table}[h]
\caption{List of species that contain 98\% of the sulfur budget at r=200 au in our \texttt{Nautilus} model.}
\label{tab:S_major_carriers}
\begin{center}
\begin{tabular}{lcc}
\hline \hline
Species & Abundance & Fraction of total (\%) \\
\hline
\multicolumn{3}{c}{Molecular cloud}\\
\hline
S & 3.9$\rm \times 10^{-6}$ & 26.0 \\
g-HS & 3.8$\rm \times 10^{-6}$ &  25.4 \\
g-\ce{H2S} & 3.3$\rm \times 10^{-6}$ & 22.2 \\
g-NS & 2.7$\rm \times 10^{-6}$ & 17.8 \\
g-S & 5.9$\rm \times 10^{-7}$ & 3.9 \\
g-OCS & 1.9$\rm \times 10^{-7}$ &  1.2 \\
CS  & 1.7$\rm \times 10^{-7}$ & 1.1 \\
\hline
\multicolumn{3}{c}{AB Aur}\\
\hline
g-\ce{H2S} & 9.2$\rm 10^{-6}$ & 61.4 \\
g-\ce{SO2} & 2.6$\rm \times 10^{-6}$ & 17.3 \\
g-\ce{H2S3} & 4.6$\rm \times 10^{-7}$ & 9.2 \\
S$\rm ^+$ & 1.0$\rm \times 10^{-6}$ & 6.9 \\
g-\ce{CS2} & 1.2$\rm \times 10^{-7}$ & 1.6 \\
S & 1.9$\rm \times 10^{-7}$& 1.3 \\
\hline
\end{tabular}
\end{center}
\label{default}
\tablefoot{The prefix g- means that the species is adsorbed in the surface of dust grains.}
\end{table}%

\subsection{Model results}
In Fig. \ref{Fig:sulfur_budget_per_height}, we show the cumulative fraction of the total sulfur abundance for the ten most abundant sulfuretted species computed by \texttt{Nautilus} at different heights for a given radius of 200 au, where the pressure scale height H is $\rm \sim$ 8 au. The prefix s means that the species are on the surface of dust grains. The lack of prefix means that the species are in gaseous form. The obtained distribution does not change much from z=0.1 au ($\rm A_v \sim$29 mag) to z$\rm <$7 au ($\rm A_v \sim$11.5 mag). At these heights, most of the sulfur is in the form of H$\rm _2$S on the surface of grains (more than 95\% of the cosmic abundance at z=0.1 au, and 65\% at z$\rm \sim$7 au).  At z$\rm \sim$8 au \ce{H2S3} and \ce{SO2} in the  surface of grains become the dominant  carriers. At z=10, the budget is dominated by \ce{SO2} on the surface of grains (79\% of the cosmic sulfur abundance).  At z=15 au, the budget becomes dominated by two gas species: S$\rm ^+$ and S contain 97\% of the sulfur abundance. Finally, at z=25 au, 99.8\% is in the form of S$\rm ^+$.

\begin{figure}[h!]
\begin{center}
 \includegraphics[width=0.5\textwidth,trim = 0mm 0mm 0mm 0mm,clip]{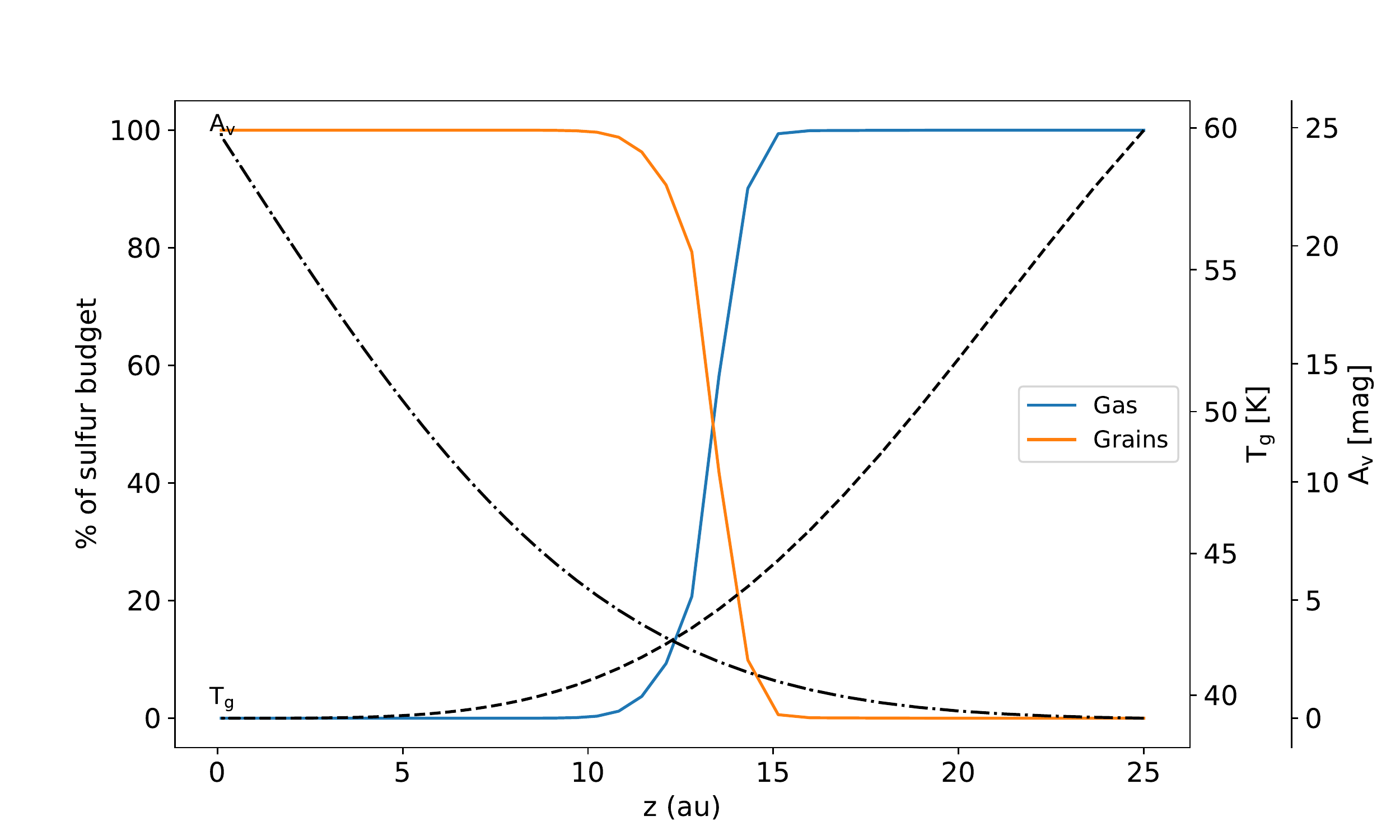}\\
  \caption{Percentage of the sulfur abundance contained in the different phases of the model as a function of height at  200 au: in the mantle of grains, in the surface of grains, and in gas phase.}
 \label{Fig:phases_vs_height}
\end{center}
\end{figure}

The trends with the height of the different phases of the model are summarized in Fig. \ref{Fig:phases_vs_height}, where we represent the percentage of the sulfur cosmic abundance contained in each of the phases that are considered by the model (sulfur species in the surface of dust grains and gas-phase species) as a function of the height over the mid-plane at r=200 au. As can be seen, before $\rm \sim$14 au the budget is dominated by surface species. Then, around 12 au, chemical species in the gas phase start to become important sulfur carriers. At z$\rm \sim$13 au, the amount of sulfur in the gas phase becomes larger than the amount of sulfur locked in the surface of grains, and, at 16 au, $\rm \sim$100\% of the sulfur budget is contained in the gas phase.

\begin{figure}[t!]
\begin{center}
 \includegraphics[width=0.5\textwidth,trim = 0mm 0mm 0mm 0mm,clip]{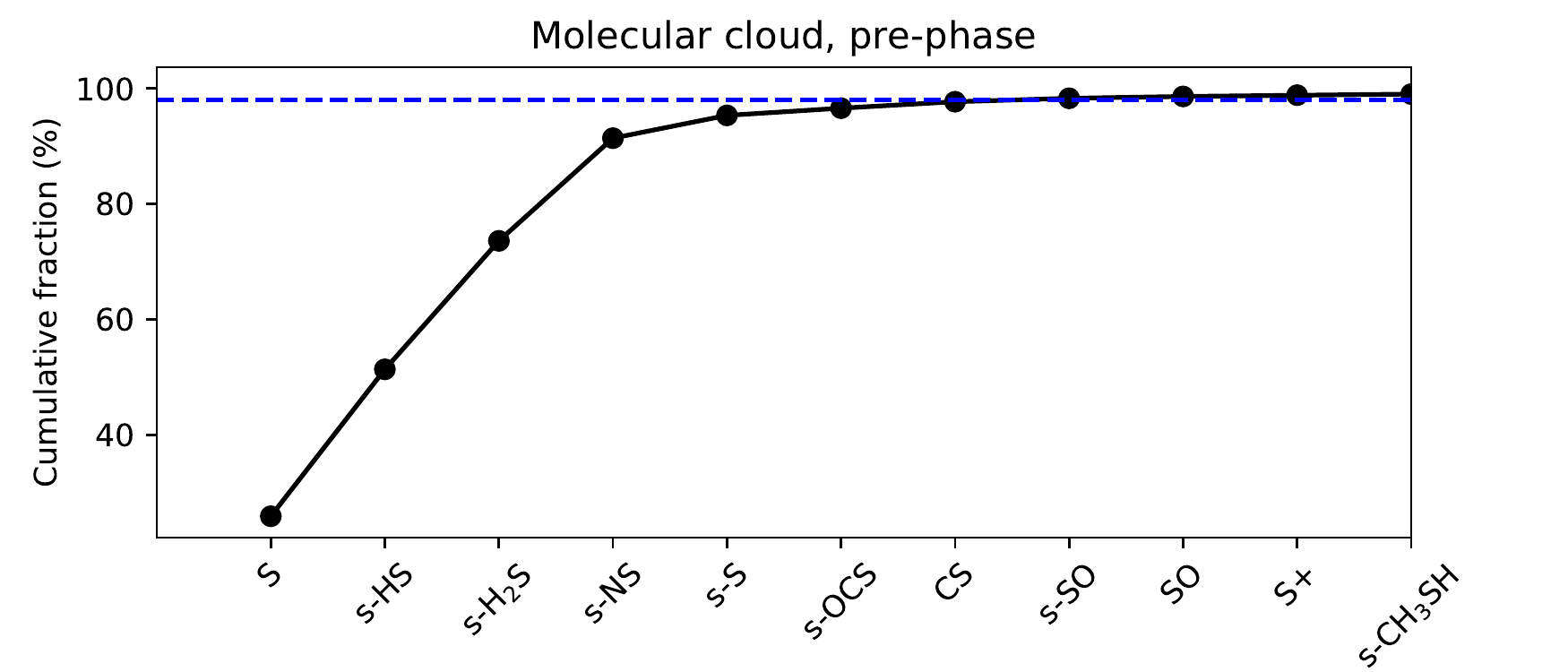}\\
 \includegraphics[width=0.5\textwidth,trim = 0mm 0mm 0mm 0mm,clip]{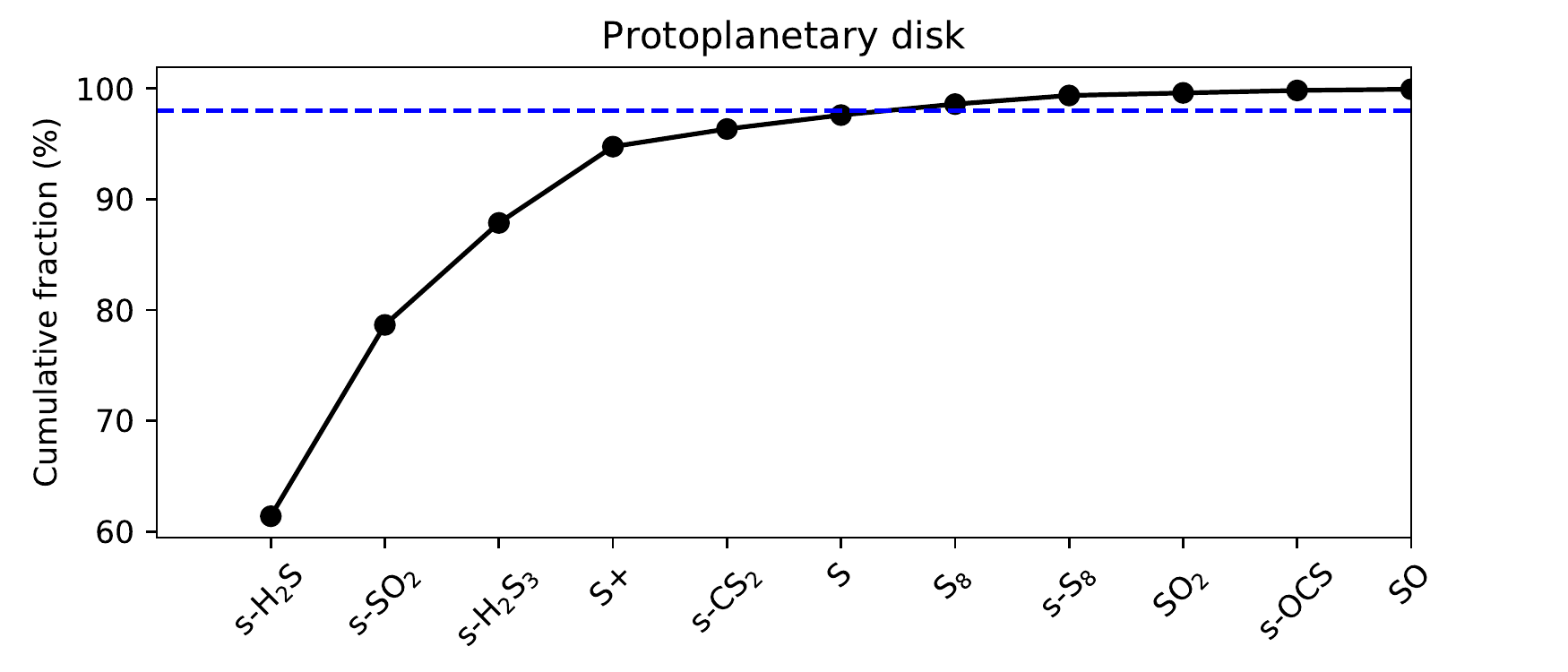}\\
  \caption{Cumulative sulfur abundance fraction of the ten main sulfur carriers from \texttt{Nautilus} models. Top: molecular cloud. Bottom: AB Aur protoplanetary disk at r=200 au. The horizontal blue dashed line depicts the position of the 98\% cumulative fraction.}
 \label{Fig:sulfur_budget}
\end{center}
\end{figure}

In Fig. \ref{Fig:sulfur_budget}, we show the total (i.e., stacking all heights) cumulative fraction for the 10 most abundant sulfuretted species for the two stages of the model: the cold molecular cloud (top) and the AB Aur protoplanetary disk (bottom). In Table \ref{tab:S_major_carriers} we list the species that contribute to 98\% of the sulfur budget in each stage of the model. As can be seen in Fig. \ref{Fig:sulfur_budget}, 98\% of the sulfur budget in the protoplanetary disk model is contained in six species originating at different heights: g-\ce{H2S}, g-\ce{SO2}, g-\ce{H2S3}, S$\rm ^+$, g-\ce{CS2}, and S. Exception made of S$\rm ^+$ and S, all of them are grain surface species. Overall, 90.5\% of the sulfur content is on the surface of grains, and 9.5\% is in the gas phase. As seen in Fig. \ref{Fig:sulfur_budget_per_height}, S$\rm ^+$ only becomes a dominant species at z$\rm \sim$13 au, where the temperature reaches 43 K, and the extinction is only 2.4 mag. \ce{H2S} is also an important sulfur carrier in the gas phase, with an abundance of 7.4$\rm \times 10^{-10}$, compared to 9.2$\rm \times 10^{-6}$ for \ce{H2S} on the surface of grains. It becomes apparent that the main sulfur carriers are \ce{H2S} and \ce{SO2}, but they are almost fully incorporated in the surface and mantle of dust grains. \ce{H2S} in the gas phase contains only $\rm \sim$0.003\% of the total sulfur budget. 

\begin{figure}[t!]
\begin{center}
 \includegraphics[width=0.5\textwidth]{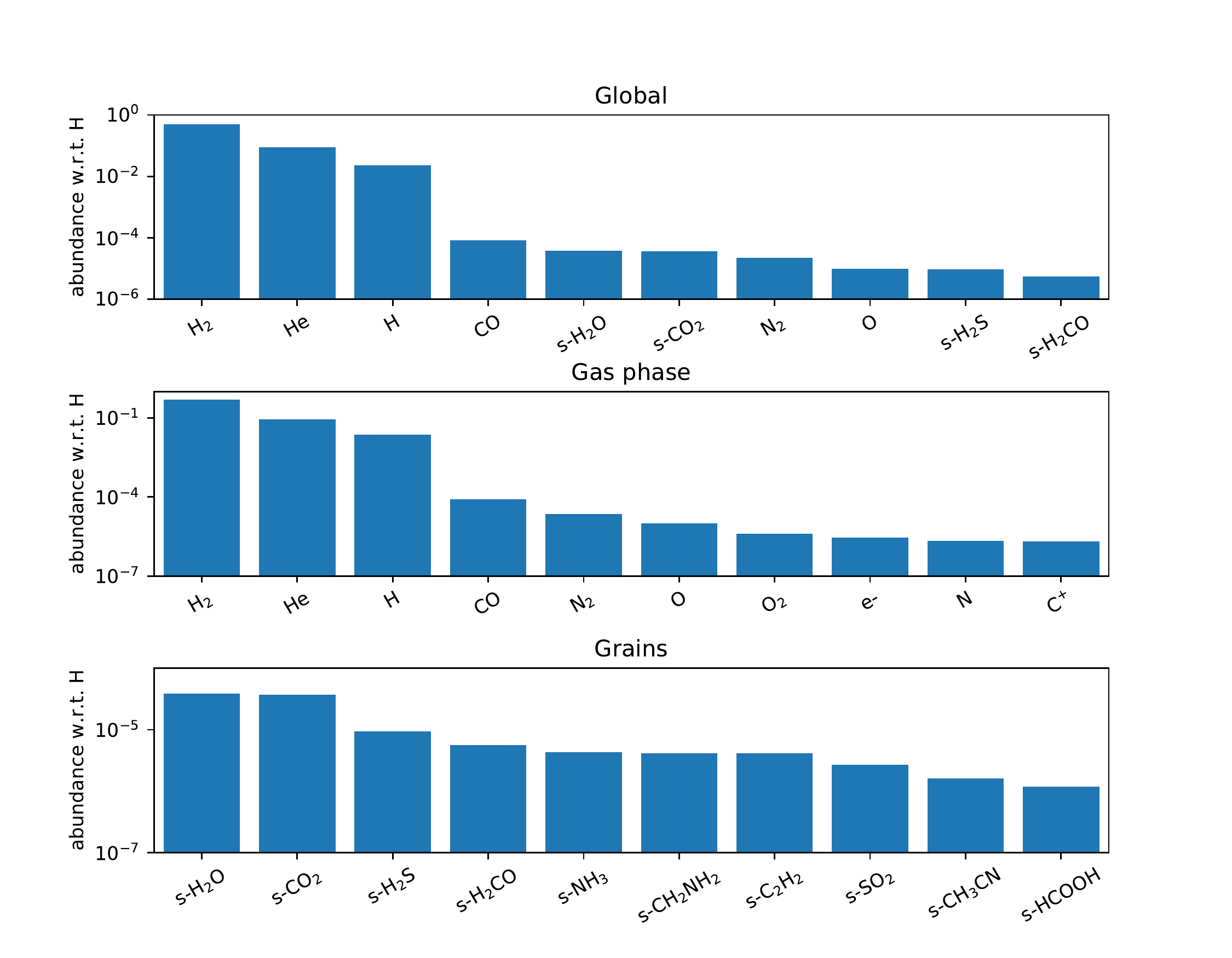}\\
  \caption{Abundances of the ten most abundant species in our \texttt{Nautilus} model for the global content (top), the gas phase (middle), the surface of grains (bottom).}
 \label{Fig:global_budget}
\end{center}
\end{figure}

In Fig. \ref{Fig:global_budget} we show the 10 most abundant species for the different phases of the model, as well as for the global case. Surface \ce{H2S} is the third most abundant species, after g-\ce{H2O} and g-CO. Furthermore, considering all phases (Fig. \ref{Fig:global_budget}, top left) g-\ce{H2S} is the ninth most abundant species in the model. This makes \ce{H2S} a relevant component of dust grains, whose abundance in different protoplanetary systems is worth modeling.

\begin{figure*}[t!]
\begin{center}
 \includegraphics[width=1.0\textwidth,trim = 0mm 0mm 0mm 0mm,clip]{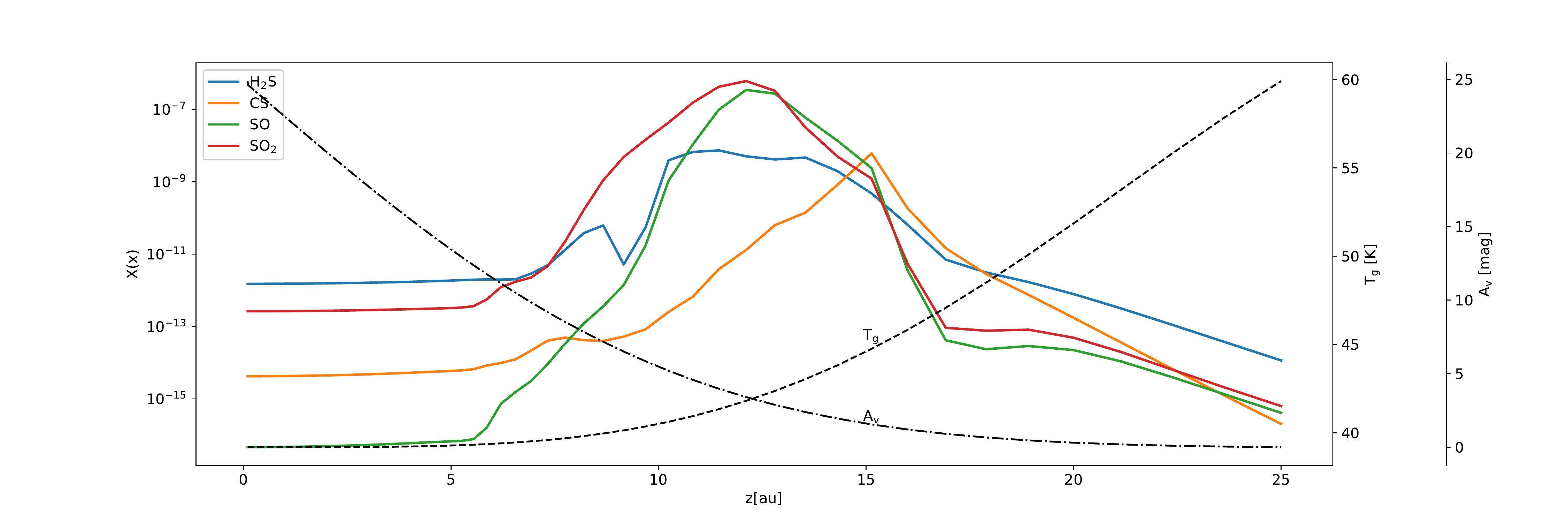}\\
  \includegraphics[width=1.0\textwidth,trim = 0mm 0mm 0mm 0mm,clip]{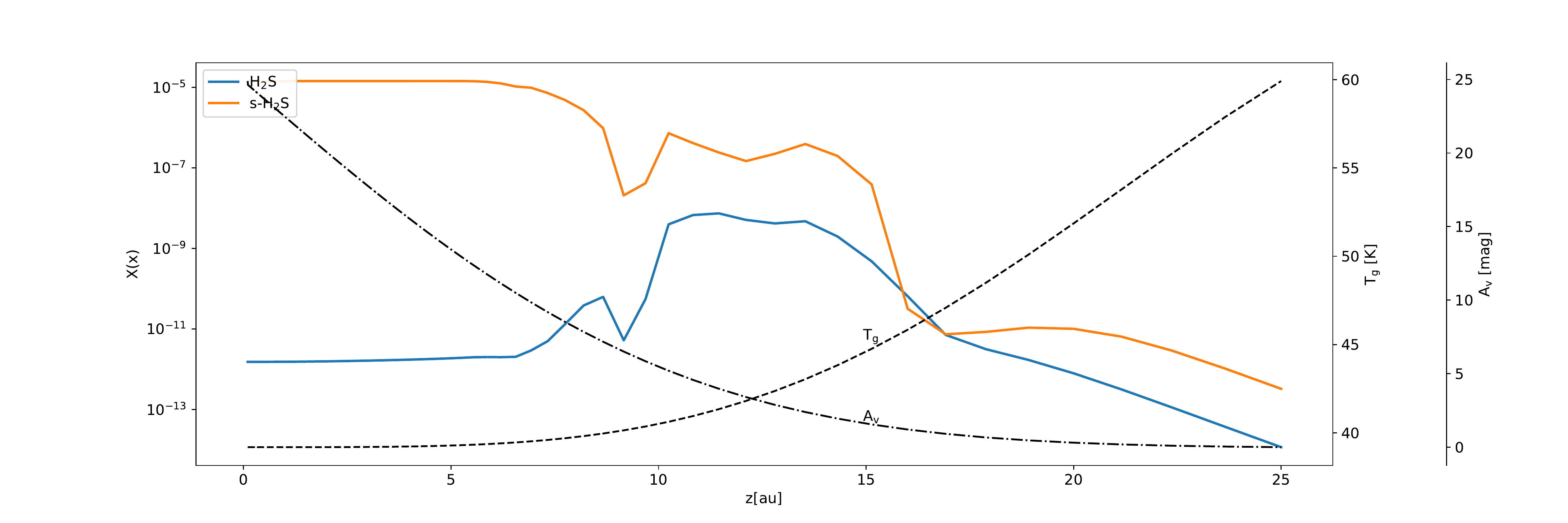}\\
  \caption{Vertical profiles at r=200 au. Top: of  important sulfur carriers in the gas phase, namely \ce{H2S}, CS, SO, and \ce{SO2}. Bottom: of \ce{H2S} in the gas phase and in the surface of grains (g-).We also include the vertical profile of $\rm A_v$ and $\rm T_g$ for comparison.}
 \label{Fig:ab_S_height}
\end{center}
\end{figure*}

In Fig. \ref{Fig:ab_S_height}, top panel, we show the vertical profile at r=200 au of a few sulfur carriers that are important for the sulfur budget in the gas phase, namely: \ce{H2S}, CS, SO, and \ce{SO2}. This figure helps us to identify at which height over the mid-plane the emission of the different species originates. Before z$\rm \sim, $ 7 au the height profiles of the four species remain mostly flat at low abundances. At 7 au, their abundances start to grow, and  SO, and \ce{SO2} peak around 13 au, while CS peaks at 15 au. The \ce{H2S} vertical profile forms a high abundance plateau that extends from $\rm \sim$10 to $\rm \sim$15 au where the abundance reaches values of a few $\rm 10^{-9}$. The plateau is preceded by a dip where the abundance goes down by three orders of magnitude. In the bottom panel of Fig.  \ref{Fig:ab_S_height} we show the same as in the top panel, but this time for the different phases of \ce{H2S}: gas phase and \ce{H2S} in the surface of grains (g-\ce{H2S}). At low heights over the mid-plane (high extinction), the dominant phase is \ce{H2S} on the surface of grains.

\subsection{Parameter impact}
\begin{figure}[h!]
\begin{center}
 \includegraphics[width=0.5\textwidth,trim = 0mm 0mm 0mm 0mm,clip]{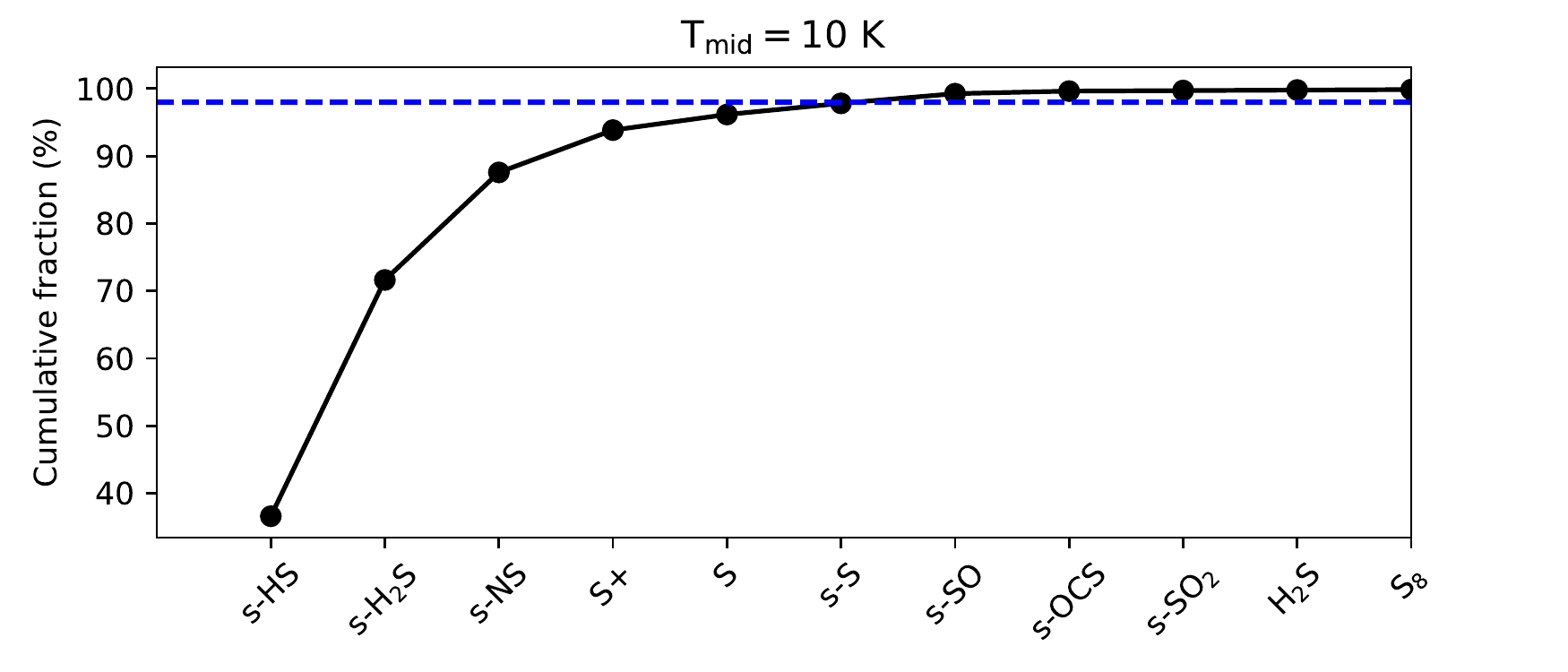}\\
 \includegraphics[width=0.5\textwidth,trim = 0mm 0mm 0mm 20mm,clip]{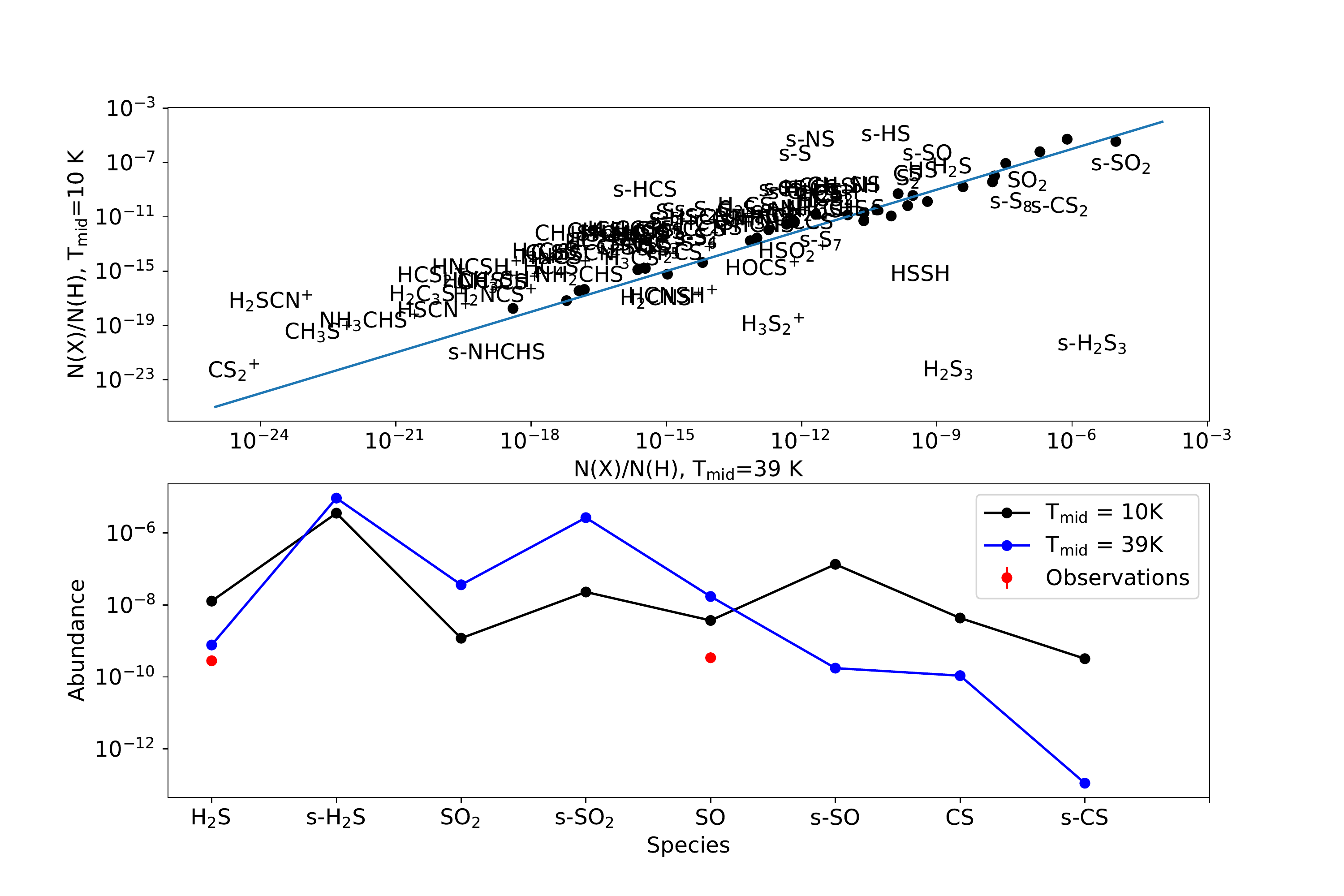}\\
  \caption{Impact of varying the mid-plane temperature on the model results. Top: cumulative sulfur abundance fraction of the ten main sulfur carriers for the \texttt{Nautilus} model with $\rm T_{mid}$=10 K. Middle: abundance of sulfuretted species computed assuming $\rm T_{mid}$=10 K versus the abundance computed assuming $\rm T_{mid}$=39 K. We show the names of the species when there is an order of magnitude difference between both models, and plot black dots otherwise. Bottom: abundance of a subset of the most relevant sulfur carriers for the two mid-plane temperatures used.}
 \label{Fig:impact_of_mid_plane_T}
\end{center}
\end{figure}

In the following subsection, we discuss the impact of changing key model parameters on the results. The first parameter whose impact we wanted to test is the temperature of the mid-plane. To that aim, we computed a model with a mid-plane temperature of 10 K, compared to 39 K assumed in the previous model (see Table \ref{tab:model_parameters}). The cumulative sulfur abundance fraction of the ten most relevant sulfur carriers from this model is shown in the top panel of Fig. \ref{Fig:impact_of_mid_plane_T}. The most relevant sulfur carrier for the model with $\rm T_{mid}=10$ K is g-HS, it is not in the list of the 10 most relevant sulfur carriers in the model with $\rm T_{mid}=39$ K. The same happens with a g-NS, which are among the most relevant sulfur carriers in the $\rm T_{mid}=10~K$ model. Species that are important sulfur carriers in the main model ($\rm T_{mid}=39~K$) are irrelevant to the sulfur budget in the $\rm T_{mid}=10~K$ model. Such species include g-\ce{SO2} and  g-\ce{H2S3}. We show in the middle of Fig. \ref{Fig:impact_of_mid_plane_T} a comparison of the abundances of all the sulfuretted species from the model with $\rm T_{mid}=10$ K versus the abundances from the model with $\rm T_{mid}=39$ K. The figure presents a large scatter, indicating that varying the temperature of the mid-plane gas has an important impact on the derived sulfur abundances. In the bottom panel of Fig. \ref{Fig:impact_of_mid_plane_T}  we show the abundances of a subset of important sulfur carriers. While the differences are small for \ce{H2S} in the surface and mantle of grains, they are large for all the other species represented. We include in the bottom panel the abundances derived from our observations assuming LTE. As can be seen, none of the models matches the two observed species, but the main model provides a good fit for \ce{H2S}. In the middle panel of Fig. \ref{Fig:impact_of_mid_plane_T} we see that some molecules show differences of several orders of magnitude. Therefore, a subset of these species can be used to get insight into the mid-plane temperature, according to models.

\begin{figure}[h!]
\begin{center}
 \includegraphics[width=0.5\textwidth,trim = 0mm 0mm 0mm 20mm,clip]{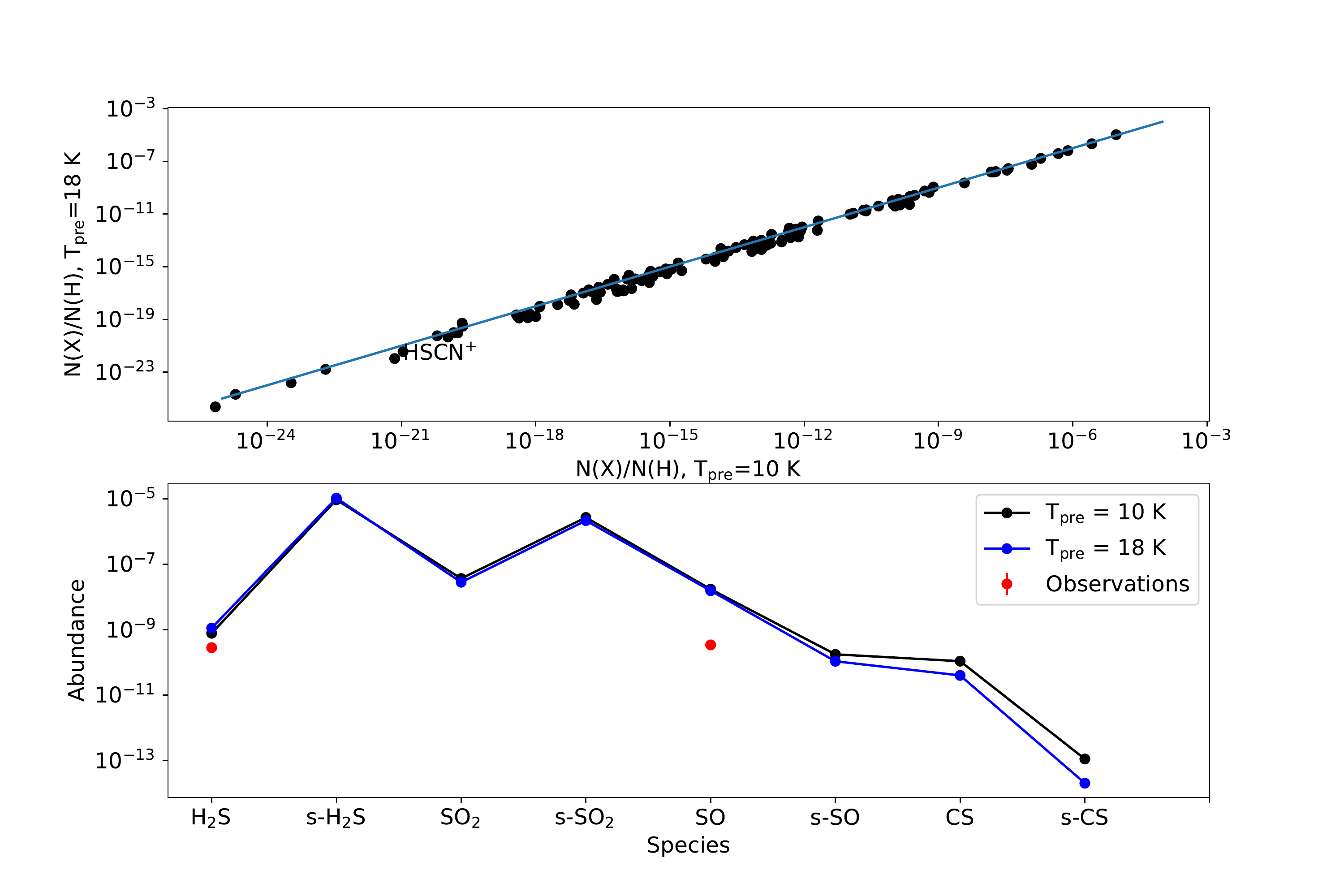}\\
  \caption{Impact of varying the prephase temperature on the model results. Top: abundance of sulfuretted species computed assuming $\rm T_{gas}$=10 K versus the abundance computed assuming $\rm T_{gas}$=18 K for the molecular cloud prephase. Bottom: abundance of a subset of the most relevant sulfur carriers for the two prephase temperatures used.}
 \label{Fig:impact_of_prephase_T}
\end{center}
\end{figure}

The second parameter that we wanted to test is the temperature of the cold cloud prephase. To that aim, we computed a model using a prephase with a temperature of 18 K. As can be seen in Fig. \ref{Fig:impact_of_prephase_T} the differences between the two models are not relevant, and a change in the prephase gas temperature will produce almost the same results, especially in the gas phase.

\begin{figure}[h!]
\begin{center}
 \includegraphics[width=0.5\textwidth,trim = 0mm 0mm 0mm 0mm,clip]{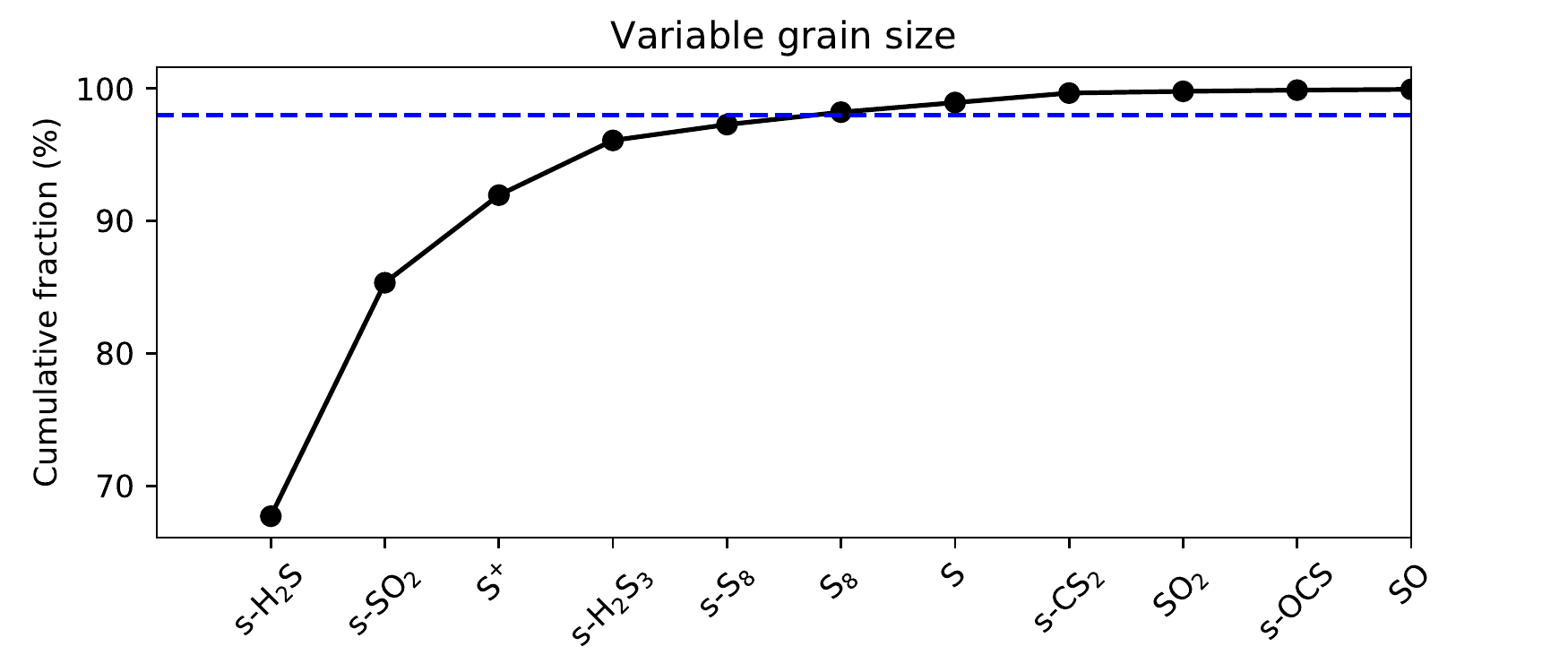}\\
 \includegraphics[width=0.5\textwidth,trim = 0mm 0mm 0mm 20mm,clip]{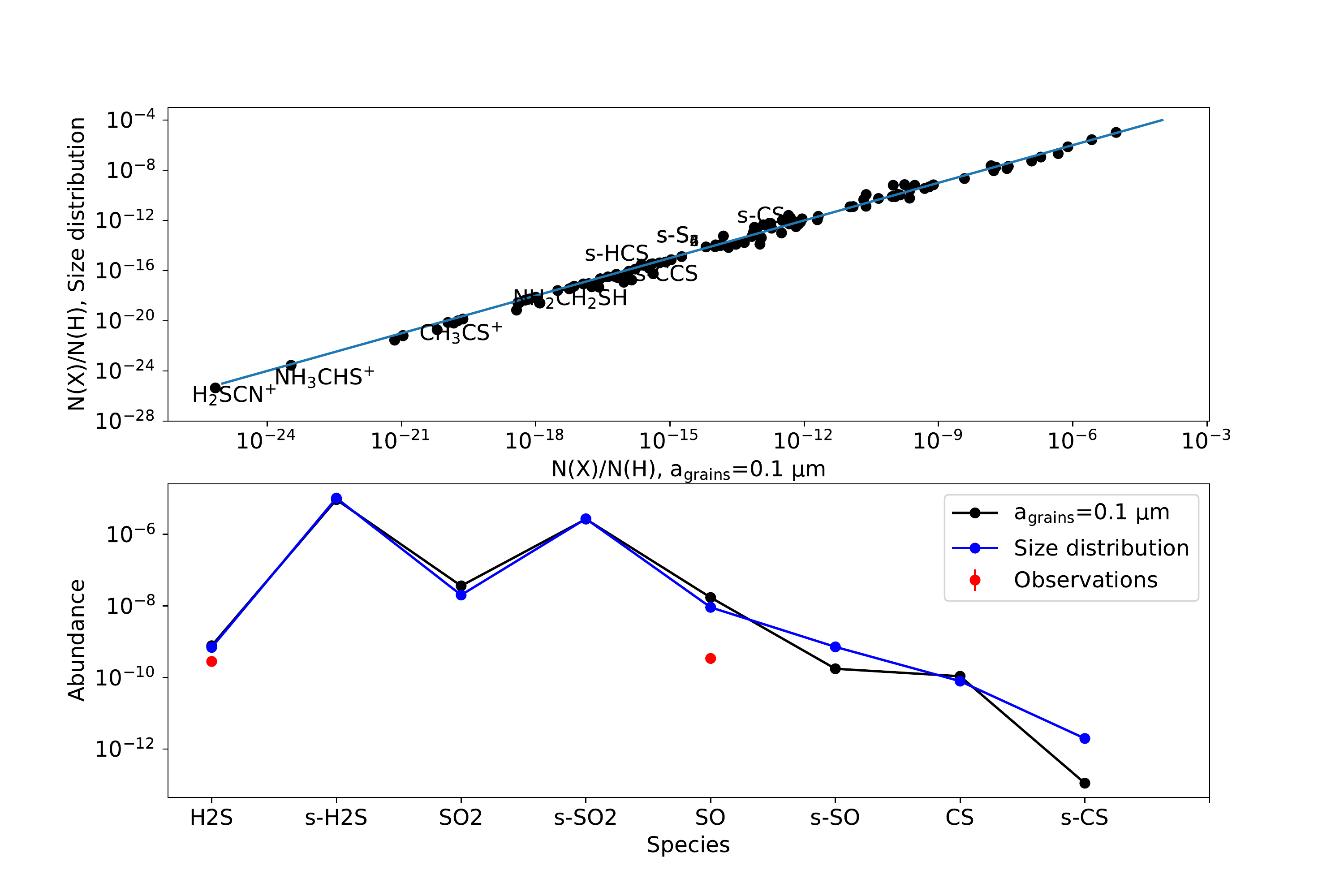}\\
  \caption{Impact of varying grain size on the model results. Top: cumulative sulfur abundance fraction of the ten main sulfur carriers for the \texttt{Nautilus} model computed assuming a grain size distribution. Middle: abundance of sulfuretted species computed assuming $\rm a_{grains}$=0.1 $\rm \mu m$ versus the abundance computed assuming a grain size distribution.  We show the names of the species when there is an order of magnitude difference between both models, and plot black dots otherwise.Bottom: abundance of a subset of the most relevant sulfur carriers for the model with fixed grain size and with a distribution of sizes.}
 \label{Fig:impact_of_grain_size}
\end{center}
\end{figure}

We also tested the impact of a varying grain size distribution. To that aim, we assumed a different grain size at each height, with 1 mm grains in the mid-plane, 0.1 $\rm \mu$ m in the atmosphere, and a logarithmic interpolation in between. We use the same dust temperature in both models. We show in Fig. \ref{Fig:impact_of_grain_size} the results from this model versus the original model results. While the difference is not as large as when changing the mid-plane temperature, the impact of grain sizes cannot be ignored. As can be seen in the top panel of Fig. \ref{Fig:impact_of_grain_size}, the species that dominate the sulfur budget are similar to those in the case of the original model (Fig. \ref{Fig:sulfur_budget}). However, although the grain size distribution affects the results, those for the gaseous species tracked in the bottom panel of Fig. \ref{Fig:impact_of_grain_size} remain mostly unchanged.

\begin{figure}[h!]
\begin{center}
 \includegraphics[width=0.5\textwidth,trim = 0mm 0mm 0mm 20mm,clip]{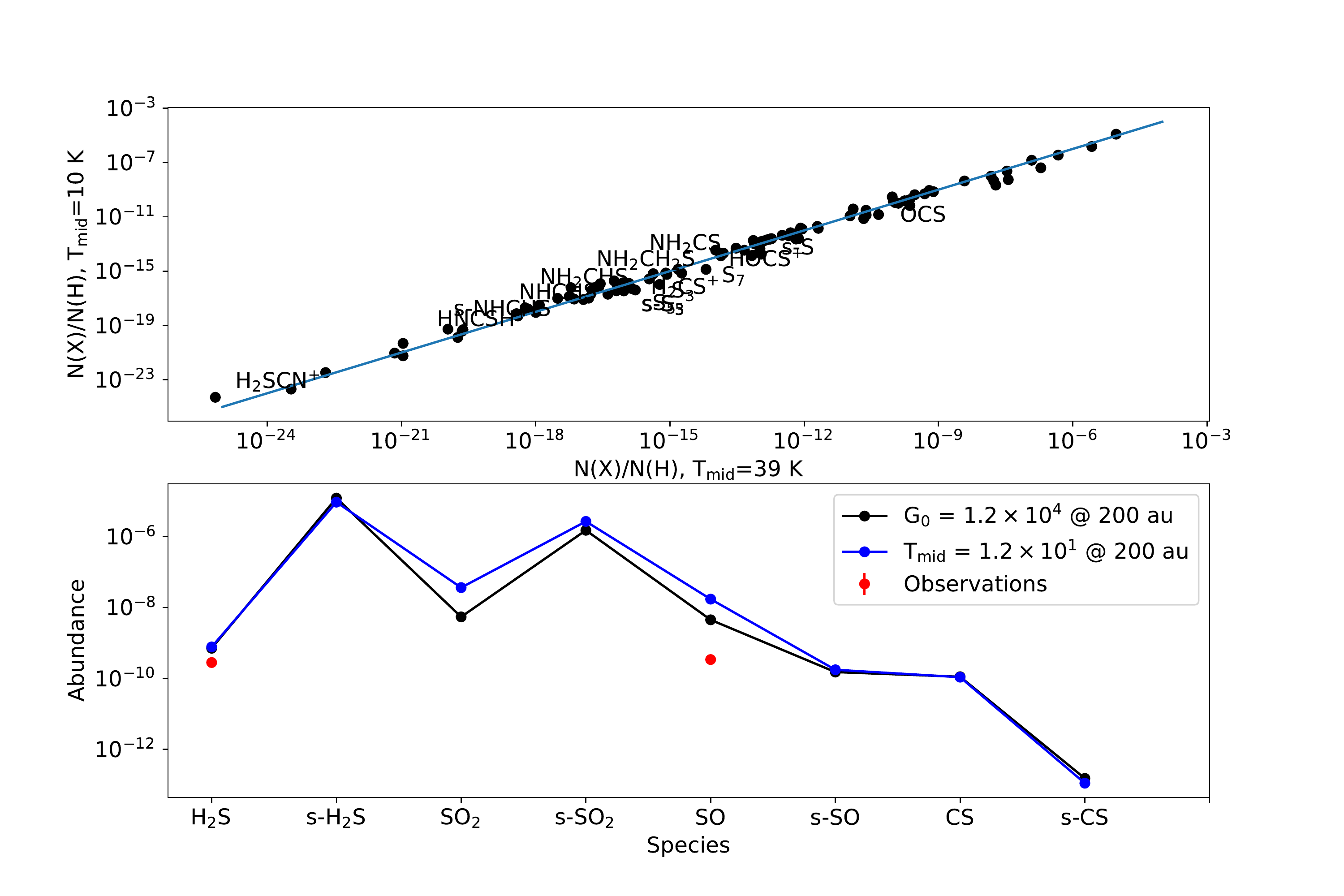}\\
  \caption{Impact of varying the UV flux on the model results. Top: cumulative sulfur abundance fraction of the ten main sulfur carriers for the \texttt{Nautilus} model computed assuming $\rm f_{UV}$=12. Middle: abundance of sulfuretted species computed assuming $\rm f_{UV}$=12 Draine unit versus the abundance computed assuming $\rm f_{UV}$=1.2$\rm \times 10^4$ Draine units. Bottom: abundance of a subset of the most relevant sulfur carriers for the model with $\rm f_{UV}$=12 Draine unit  and with $\rm f_{UV}$=1.2$\rm \times 10^4$ Draine units.}
 \label{Fig:impact_of_fUV}
\end{center}
\end{figure}

Finally, we tested the impact of the UV field by computing a model with a meager value of the scale factor of the UV flux, $\rm f_{UV}=12$, compared to $\rm f_{UV}=12\times 10^{4}$ in the main model (see Table \ref{tab:model_parameters}). We show in Fig. \ref{Fig:impact_of_fUV} a comparison between the two models. The impact on the abundances is small, and only a few species show a difference larger than one order of magnitude. One of those species is S$\rm ^+$, which drops by more than one order of magnitude when a low UV flux is assumed.

\begin{table*}[t]
\caption{Comparison of abundances in different environments}
\label{tab:ab_comparison}
\begin{center}
\begin{tabular}{cccccc}
\hline \hline
Species & AB Aur & GG Tau & Horsehead PDR & TMC 1-CP & Orion KL \\
\hline
\ce{H2S} & 2.4$\rm \times 10^{-10}$ & 1.8$\rm \times 10^{-10}$$\rm ^*$ & 3.3$\rm \times 10^{-10}$& $\rm 1.1\times 10^{-9}$ & 1.5$\rm \times 10^{-6}$  \\
SO          & 3.2$\rm \times 10^{-9}$ & --                                                   & 3.9$\rm \times 10^{-10}$ & $\rm 1.0\times 10^{-9}$ & 8$\rm \times 10^{-8}$\\
CS          & 2.6$\rm \times 10^{-10}$$\rm ^*$ & 3.1$\rm \times 10^{-10}$$\rm ^*$ & 4.5$\rm \times 10^{-10}$ & $\rm 1.3\times 10^{-8}$ & 7$\rm \times 10^{-8}$ \\ 
\hline
S budget & $\rm >9.2\times 10^{-10}$ & $\rm >4.9\times 10^{-10}$ & $\rm >1.2\times 10^{-9}$ & $\rm >1.5\times 10^{-8}$ & $\rm >1.7\times 10^{-6}$ \\ 
S budget$^\dagger$ & -- & -- & $\rm 1.5\times 10^{-9}$ & -- & $\rm1.9\times 10^{-6}$ \\
\hline
\end{tabular}
\end{center}
\label{default}
\tablefoot{($\rm ^*$): Assuming $\rm ^{12}CO/^{13}CO=60$ and $\rm X(^{12}CO)=10^{-4}$. ($\rm ^\dagger$): including other S-bearing species. References: GG Tau:  \cite{Phuong2018}; Horsehead PDR: \cite{Riviere2019b}; TMC 1-CP:  \citet{RodriguezBaras2021, Navarro2020}; Orion-KL: \citet{Tercero2010, Esplugues2014, Crockett2014}.} 
\end{table*}%

\section{Discussion}\label{Sect:discussion}
The detection of \ce{H2S} emission in a ring around AB Aur is the second resolved observation of \ce{H2S} in a protoplanetary disk, after the detection toward GG Tau by \cite{Phuong2018}, and the sixth detection after the four single dish detections in Taurus protoplanetary disk \citep{Riviere2021}. 

Assuming LTE we derived a mean column density of   (1.9$\rm \pm$0.4)$\rm \times 10^{13}~cm^{-2}$ and a mean abundance with respect to $\rm ^{13}$CO of  (1.5$\rm \pm$0.4)$\rm \times 10^{-4}$. By assuming $\rm ^{12}CO$/$\rm ^{13}CO$=60, and X($\rm ^{12}CO$)=10$\rm ^{-4}$, we converted the abundance with respect
to $\rm ^{13}CO$ to an abundance with respect
to H$\rm _2$, resulting in a mean value of  (2.4$\rm \pm$0.6)$\rm \times 10^{-10}$. The region surrounding the dust trap depicts the maximum \ce{H2S} abundance in the disk. Our \texttt{Nautilus} 1D model from Sect. \ref{sect:mod_sulfur_budget} predicts N(\ce{H2S})=3.5$\rm \times 10^{13}$ (X(\ce{H2S})=7.5$\rm \times 10^{-10}$), about two times larger than the value that we estimate assuming LTE. Using the same methodology, we derived the SO column density from our  SO 5$\rm _6$-4$\rm _5$ NOEMA observations from \cite{Riviere2020}, resulting in  a mean SO column density of 1.0$\rm \times 10^{14}~cm^{-2}$, resulting in a mean abundance of 3.2$\rm \times 10^{-9}$, $\rm \sim$2.4 times smaller than the value predicted by our \texttt{Nautilus} model, N(SO)=7.8$\rm \times 10^{14}$ (X(\ce{SO})=9.8$\rm \times 10^{-9}$).  Recently, CCS was detected toward GG Tau \citep{Phuong2021}. The authors could not simultaneously reproduce the observed column density of CCS and other sulfuretted species, supporting the idea that astrochemical sulfur networks are incomplete. The complexity of simultaneously predicting the abundance of important sulfur carriers such as CS, SO, SO$\rm _2$, and H$\rm _2$S is a known issue of sulfur chemistry networks in astrochemical models \citep{Navarro2020}. Sulfur astrochemical networks have improved over the last years \citep{LeGal2017,Vidal2017,Navarro2020}, with new formation and destruction routes. Yet, progress is needed to solve the issue. However, the goal of the model is not to reproduce the observed values, but rather to get insight into the main sulfur carriers. Furthermore, considering the uncertainties in the model, related to the physical structure and the details of the chemical network, we consider that the model results in a good fit of the \ce{H2S line}. 

From the computed column density map we derived a mean \ce{H2S} abundance with respect to $\rm ^{13}CO$ of $\rm (1.5\pm0.4)\times 10^{-4}$. Assuming a $\rm ^{12}CO$ abundance of 10$\rm ^{-4}$ and  $\rm ^{12}CO$/$\rm ^{13}CO$=60,  we derived a mean abundance with respect
to H nuclei X(\ce{H2S})=(2.4$\rm \pm$0.6)$\rm \times 10^{-10}$. To help us with the discussion we also derived abundances for SO, the other sulfuretted species that we observed toward AB Aur \citep{Riviere2020}. The mean SO abundance with respect to $\rm ^{13}CO$ is $\rm (1.9\pm0.2)\times 10^{-3}$, resulting in an abundance with respect to H nuclei  X(\ce{SO})=(3.1$\rm \pm$2.3)$\rm \times 10^{-9}$. 

In the following, we compare the \ce{H2S} abundance derived in AB Aur with that derived in different environments. \cite{Phuong2018} used \texttt{Nautilus} to derive a column density for the disk around the M0 star GG Tau N(\ce{H2S})=3.4$\rm \times 10^{13}~cm^{-2}$, close to our estimate for AB Aur. \cite{Riviere2021} detected \ce{H2S} in four young stellar objects in Taurus, with column densities in the range 3.2$\rm \times 10^{12}~cm^{-2}$ to 1.9$\rm \times 10^{13}~cm^{-2}$. This set of values was computed for a disk with an outer radius of 500 au. However, this size is larger than the \ce{H2S} disk observed toward AB Aur. If an outer radius of 250 au is assumed, the computed column densities are in the range $\rm 10^{13}~cm^{-2}$ to 7$\rm \times 10^{13}~cm^{-2}$, close to the value derived for AB Aur. The difference in spectral type does not impact the individual column densities. Since  \ce{H2S} has been detected only in a few protoplanetary disks, no statistical trend can be derived, but the proximity between observed abundances is interesting.

The abundance derived for AB Aur is four orders of magnitude lower than that found in the Orion KL hot core  \citep[1.5 $\rm \times 10^{-6} $, ][]{Esplugues2014}, but similar to that found in two PDRs, the Horsehead nebula \citep[3.1$\rm \times 10^{-10}$,][]{Riviere2019} and the  Orion Bar \citep[4.0$\rm \times 10^{-10}$,][]{Leurini2006}. This is particularly interesting because \ce{H2S} presents almost the same abundance in very different PDRs (the Horsehead with a radiation intensity of G$\rm _0 \sim 10^2$ and the Orion Bar, a much more extreme PDR with G$\rm _0$=$\rm 10^4$. Overall, the chemical abundances in PDRs are governed by the ratio of the UV radiation field and the density ($\rm G_0/n_H$). In the case of H2S, this species is mainly formed on grain surfaces and later released into the gas phase through photodesorption in regions dominated by high UV radiation  \citep{JimenezEscobar2011, JimenezEscobar2014, Fuente2017}. The high UV radiation is also responsible for the destruction of \ce{H2S} through dissociation processes. The interplay between formation and destruction mechanisms makes the position of the \ce{H2S} abundance peak roughly independent of G$\rm _0$. This is partially confirmed by our experiment with a low-UV model, which results in very close \ce{H2S} abundances: 6.9$\rm \times 10^{-10}$ for the model with a low UV field versus 7.5$\rm \times 10^{-10}$ for the model with high UV field.

Regarding density, it has a crucial role in the \ce{H2S} abundance, since both the formation of solid \ce{H2S} on the surface of grains and its desorption depend on the density of the region. In particular, solid \ce{H2S} is mainly formed through recombination of S$\rm ^+$ ions (formed as the UV photons penetrate the gas ionizing S atoms) with negatively charged grains followed by desorption. Adsorption of S$\rm ^+$ is favored in dense regions by the increased collision rates between gas-phase species and grains \citep{Gail1975, Bel1989, Ruffle1999, Druard2012}. Once that solid \ce{H2S} is formed, the \ce{H2S} photodesorption rates depend on the grain size distribution, which in turn depends on $\rm n_H$. This implies that, for a given radiation field, the \ce{H2S} abundance peak will be shifted toward the densest part of a PDR as theoretically found by \cite{Goicoechea2021}.

The huge change in the abundance of \ce{H2S} in the gas phase from hot cores to Class II disks and PDRs points to \ce{H2S} being an important sulfur carrier in the icy surface of dust grains, such that it is almost fully evaporated in hot cores, thus increasing the fraction of gas-phase \ce{H2S}. This is confirmed by our \texttt{Nautilus} model from Sect. \ref{sect:mod_sulfur_budget}, where only $\rm \sim$ 0.0001\% of \ce{H2S} is in gas phase. In this model, a large fraction ($\rm \sim$61\%) of the cosmic sulfur abundance is locked in the surface and mantle of grains in the form of \ce{H2S}, and another 17\% is also locked in the surface and mantle of grains in the form of \ce{SO2}. The low sulfur abundances observed in AB Aur and other protoplanetary disks \citep{LeGal2019} can indeed be explained if sulfur is locked in the ice surface of dust grains, similar to what happens in dense cores \citep{Millar1990, Ruffle1999, Vidal2017, Laas2019}. \cite{Laas2019}, using a chemical network with 860 species, showed that most sulfur is locked on dust grains and that grain chemistry could account for the depleted sulfur in molecular clouds. Our 1D protoplanetary disk model shows a similar result, with most of the sulfur locked in the surface of dust grains in the form of \ce{H2S}, \ce{SO2} and \ce{H2S3} at high extinctions, and dominated by S$\rm ^+$ at low extinctions. 

We note that our results for the sulfur budget depend on the chemical network used, as well as on the details of the model. For instance, in the molecular cloud model by \cite{Laas2019} the amount of \ce{H2S} in ices is negligible for most of the cloud's life. The chemical network used in \cite{Laas2019} is different from the one we used in the present paper, and their model does not include a mantle phase, which results in very different abundances of \ce{H2S} on the surface of grains. We highlight, however, that \ce{H2S} is the most abundant sulfuretted species in comets \citep{Bockelee2017}, in agreement with our results.

\cite{Cazaux2022} find that, under efficient self-shielding conditions, \ce{H2S} survives on the surface of grains as the ice below 100 K, and is largely depleted from the grains at 150 K. However, if no self-shielding is assumed, most \ce{H2S} is transformed, under the influence of UV radiation, into S$\rm _x$ sulfur chains. Our \texttt{Nautilus} model predicts that gaseous S$\rm _8$ and that the surface of grains is the seventh and eighth major sulfur carrier.

\section{Summary and conclusions}\label{Sect:summary}
In this study we present resolved observations of \ce{H2S} $\rm 1_{01}-1_{10}$ in AB Aur. This is only the second time that \ce{H2S} has been resolved in a protoplanetary disk, and the sixth global detection. The main results from our study can be summarized as follows:\\

\noindent
1. We detected o-\ce{H2S} $\rm 1_{01}-1_{10}$ in a ring extending from $\rm \sim$0.67$\rm \arcsec$ ($\rm \sim$109 au) to 1.69$\rm \arcsec$ ($\rm \sim$275 au). We observed strong azimuthal asymmetries. The position of the peak coincides with the position of the continuum and C$\rm ^{18}$O emission peaks. The radial profile of the emission overlaps with that of \ce{H2CO}.\\

\noindent
2. Assuming LTE, we estimated a mean column density of $\rm (1.9\pm0.4)\times 10^{13}~cm^{-2}$. Making simple assumptions, we translated this column density into an abundance of (2.4$\rm \pm$0.6)$\rm \times 10^{-10}$.\\

\noindent
3. A \texttt{Nautilus} 1D model shows that most (99.99\%) of the \ce{H2S} is locked in the surface of dust grains. The model also shows that they are the main sulfur carriers. Our model further shows that $\rm \sim$90.5\% of the sulfur is locked on the surface of grains, and only 9.5\% of it is available in the gas phase.

Our study of the sulfur budget in the protoplanetary disk surrounding AB Aur points to \ce{H2S} being the most important sulfur carrier on the surface of grains. Furthermore, it is the third most abundant species on the surface of grains, after \ce{H2O} and \ce{CO2}. Our results probe that \ce{H2S} observations are an essential diagnostic to determine the sulfur depletion in protoplanterary disks. Observations of \ce{H2S} and other sulfuretted species toward more young stellar objects are needed to understand sulfur depletion and put our observations of AB Aur in context. 

\begin{acknowledgements}
PRM and AF thank the Spanish MINECO for funding support from PID2019-106235GB-I00.
\end{acknowledgements}

 \bibliographystyle{aa} 
\bibliography{biblio}

\end{document}